\documentclass{article}

\usepackage{arxiv}

\usepackage[utf8]{inputenc} 
\usepackage[T1]{fontenc}    
\usepackage{hyperref}       
\usepackage{url}            
\usepackage{booktabs}       
\usepackage{amsfonts}       
\usepackage{nicefrac}       
\usepackage{microtype}      
\usepackage{lipsum}
\usepackage{graphicx}

\usepackage{setspace}
\usepackage{bbm}
\usepackage{makecell}
\usepackage{dsfont}
\usepackage{enumitem}

\usepackage{pdflscape}
\usepackage{multirow}
\usepackage{threeparttable}
\usepackage{rotating}
\usepackage{amsmath}
\usepackage{bm}

\usepackage[superscript,biblabel,sort]{cite}

\title{G-computation for increasing performances of clinical trials with individual randomization and binary response}

\author{
  Joe de Keizer \\
  CIC INSERM 1402 \\
  University and University hospital of Poitiers \\
  France \\
   \And
  Rémi Lenain\\
  Service de Néphrologie\\
  University Hospital of Lille\\
  France\\
  \And
  Raphaël Porcher\\
  Centre de Recherche Epidémiologie et Statistiques\\
  UMR1153 Inserm - Université Paris Cité\\
  France\\
   \And
  Sarah Zohar\\
  HeKA,  Inria\\
  France\\
    \And
  Arthur Chatton \\
  Médecine Sociale et Préventive \\
  Faculté de Médecine, Université Laval \\
  Canada \\
  \And
  Yohann Foucher \\
  CIC INSERM 1402 \\
  University and University hospital of Poitiers \\
  \texttt{yohann.foucher@univ-poitiers.fr} \\
}

\begin{document}

\maketitle

\begin{abstract}
In a clinical trial, the random allocation aims to balance prognostic factors between arms, preventing true confounders. However, residual differences due to chance may introduce near-confounders. Adjusting on prognostic factors is therefore recommended, especially because the related increase of the power. In this paper, we hypothesized that G-computation associated with machine learning could be a suitable method for randomized clinical trials even with small sample sizes. It allows for flexible estimation of the outcome model, even when the covariates' relationships with outcomes are complex. Through simulations, penalized regressions (Lasso, Elasticnet) and algorithm-based methods (neural network, support vector machine, super learner) were compared. Penalized regressions reduced variance but may introduce a slight increase in bias. The associated reductions in sample size ranged from 17\% to 54\%. In contrast, algorithm-based methods, while effective for larger and more complex data structures, underestimated the standard deviation, especially with small sample sizes. In conclusion, G-computation with penalized models, particularly Elasticnet with splines when appropriate, represents a relevant approach for increasing the power of RCTs and accounting for potential near-confounders.
\end{abstract}

\keywords{Causal inference \and  G-computation \and Machine learning \and Randomized clinical trial  \and Type-II error}

\section{Introduction}\label{sec1}

In a clinical trial with individual randomization (RCT),  the unpredictable treatment allocation aims to balance the prognostic factors between the compared arms. By definition, this process prevents true confounders. However, residual differences due to chance may persist. Similarly to the concept of near-violation of positivity, \cite{leger_causal_2022} where violation is due to randomness rather than a systematic bias, one can define near-confounders as unbalanced prognostic factors due to sample-to-sample fluctuations. \cite{austin_statistical_2010} Therefore, methods for adjustment are increasingly used in RCTs, also because of the related increase in the statistical power, as they reduce the variance of the estimated effects. \cite{lingsma_covariate_2010, williamson_variance_2014, senn_statistical_2021} Today, the consolidated standards of reporting trials (CONSORT), the European Medicines Agency (EMA), and the guidelines of the Food and Drug Administration (FDA) all recommend adjusted analyses.\cite{noauthor_consort_2010,center_for_drug_evaluation_and_research_fda_adjusting_2023,european_medical_agency_guideline_2015}

One option is to use multiple regressions, which provide estimates of  conditional  treatment effects,  while RCTs often aim to  estimate marginal treatment effects. \cite{center_for_drug_evaluation_and_research_fda_adjusting_2023} For some measures, such as the odds ratio, they differ due to non-collapsibility.\cite{greenland_interpretation_1987} Estimating marginal effects is important as it provides average treatment effects across the entire population. The use of causal inference methods such as propensity scores based on the regression of the treatment allocation, \cite{williamson_variance_2014, zeng_propensity_2021}  G-computation (GC) based on the outcome regression, \cite{snowden_implementation_2011} or even doubly robust approaches (based on both the allocation and outcome regressions), \cite{lendle_targeted_2013} are necessary for targeting marginal estimands. In contrast, a significant proportion of the studies which reported the best performances of adjusted results from RCTs were based on conditional multiple regressions. Given those differences, the comparison of conditional and marginal effects make little sense. \cite{daniel_making_2021}

Methods for estimating marginal effects have been mainly compared in the context of real world data (RWD), while RCTs have different particularities. For instance, doubly robust approaches have been favored, in particular because they offer the advantage of a protection against misspecification: only one of the two regressions needs to be correctly specified. \cite{funk_doubly_2011} But this advantage may be discussed in the context of RCT. Firstly,  the model of the treatment allocation should be simple with a low risk of misspecification, close to the expected binomial distribution with no causal factors of treatment allocation, as near-confounders are due to sample-to-sample fluctuations. Secondly, it is counter-intuitive to model the allocation treatments in a RCT, while predictive models of the outcomes are routinely performed from RCT.  Thirdly, GC was reported with the higher power in several simulation-based studies. \cite{colson_optimizing_2016, chatton_g-computation_2020} Thus, GC may offer an interesting solution to adjust for covariates in RCT.

Recently, Tackney et al.\cite{tackney_comparison_2023} proposed a simulation-based study to investigate the impact of model misspecification on the performances of the previous methods. In the RCT context, when there is a non-linear interaction of treatment with a skewed covariate and small sample size, the authors reported that all adjustment methods suffered from bias. Moreover, variances were underestimated. These results suggest that it is important to use methods for estimating marginal effects when the sample size is sufficiently large ($n>200$). Nevertheless, regarding the previous literature in the context of RWD,\cite{le_borgne_g-computation_2021} we hypothesized that GC associated with machine learning (ML), may allow for a flexible estimation of the outcome model and could be a suitable method for RCTs with small sample sizes. In this paper, we investigated by simulations the performances of the GC according to various ML algorithms in the context of RCTs with a 1:1 allocation ratio, a superiority design and binary outcomes. We also illustrated their usefulness by reanalyzing two individual randomized trials.

\section{Methods}\label{sec2}

\subsection{G-computation}

\subsubsection{Counterfactual predictions}

Let $(Y,A,X)$  be the binary outcome ($Y = 1$ for events and 0 otherwise), the randomized treatment ($A = 1$ for treated individuals and 0 for controls),  and $X$ the set of baseline covariates.  Consider a RCT with $n$ individuals of the observed triplet ($y_i,  a_i,  x_i$), for $i=1,...,n$. Consider two potential outcomes,  denoted as $Y(1)$  and $Y(0)$,  which correspond to the outcome observed in the hypothetical scenarios of experimental and control treatments, respectively. To estimate the marginal effect, i.e. the difference in the average of the outcomes if individuals had been randomized in the experimental arm versus the control one, the GC consists of two steps.\cite{rubin_estimating_1974} First, one fits the outcome model, then defined as $Q(Y|A,X)$, from all subjects included in the RCT. Then, one predicts the potential outcomes for each subject $i$ as if they were included in the two arms. This is achieved by holding their respective covariates at the observed values $x_i$, while setting $a_i=k$: $\hat{Y}_i(k) = \hat{Q}(Y| k, x_i)$,  for $k=0,1$. We then note the marginal proportion of counterfactual events in the arm $k$ as $\hat{\pi}_k = n^{-1}  \sum_{i=1}^n  \hat{Y}_i(k) $.  We will study the difference in the marginal proportion, $\hat{\Delta} = \hat{\pi}_1 -  \hat{\pi}_0$,  and  the related odds-ratio,  $  mOR =  \dfrac{ \hat{\pi}_1 (1-\hat{\pi}_1 )}{ \hat{\pi}_0 (1-\hat{\pi}_0}$.

\subsubsection{Machine learning techniques}

We considered several models and algorithms (hereafter referred to as learners) to fit $Q(Y| A, X)$,  as we previously tested in the RWD context.\cite{le_borgne_g-computation_2021} We restricted the study to learners for which the treatment could be fixed. We estimated the tuning parameters by maximizing the average area under the receiver operating characteristic (ROC) curve (AUC) of a 20-fold cross-validation. We performed all the analyses using R version 4.3.0, using the \texttt{caret} package with a tuning grid of length equal to 20. Below, we provide a brief overview. 

\begin{itemize}

\item \textit{Lasso logistic regression.} L1 regularization allows for the selection of the predictors. To establish a flexible model, we considered all possible interactions between the treatment arm $A$ and the covariates $X$.  Additionally,  we used B-splines for the continuous covariates.  The \texttt{glmnet} package was used. The penalization of the L1 norm was the only tuning parameter.

\item \textit{Elasticnet logistic regression.} This approach mirrors the logistic regression mentioned earlier but incorporates both the L1 and the L2 regularizations.

\item \textit{Neural network.}  We chose one hidden layer, which represents one of the most common network architectures. Its size constitutes the tuning parameter. The \texttt{nnet}  package was used.

\item \textit{Support vector machine.} To relax the linear assumption,  we opted for the radial basis function kernel.  The \texttt{svmRadial} function of the \texttt{kernlab} package was used.  It  requires two tuning parameters: the cost penalty of miss-classification and the flexibility of the classification.

\item \textit{Super learner.} We also tested a super learner with the ensemble of the previous ML techniques.  Super learner consisted in a weighting average of the learner-specific predictions by using a weighted linear predictor.  In alignment with our previous choices, \cite{le_borgne_g-computation_2021} we estimated the weights by maximizing the average AUC through a 20-fold cross-validation.  We used the \texttt{SuperLearner} package.

\end{itemize}

\subsubsection{Covariates selection}

One of the key distinctions between prediction and causality lies in the selection of covariates. Establishing a robust understanding of the causal relationship structure plays a crucial role in enabling accurate causal inference.  For RWD-based analyses, this comprehension necessitates the careful exclusion of mediators,  colliders,  and instrumental variables from the modeling process. In the present context of RCT, the causal relationships are much simpler: there is no causal factor of the treatment allocation.  As a result,  we have to consider the prognostic factors of the outcome measured before randomization, except for the consequences of the treatment allocation. 

\subsubsection{Variance estimation}

We estimated the standard errors and confidence intervals by non-parametric bootstrapping.  To balance computational efficiency,  we estimated the tuning parameters using the entire data set and then applied these values to subsequent bootstrap samples. 
Given that not all of the previously mentioned methods are appropriate for small sample sizes,\cite{riley_penalization_2021} we used a bootstrap cross-validation procedure to address the potential over-fitting. This involved training the learners on the bootstrap samples and estimating the marginal treatment effect for individuals not in the learning samples.
Note that this methodology associated with ML is interesting since the construction of the outcome model can be considered in the variance estimation (i.e., valid post-selection inference).  We performed 500 bootstrap samples for the simulations and 1,000 bootstrap samples  for the applications.

\subsection{Simulations}

The R codes for the simulations are available at https://github.com/chupverse

\subsubsection{Data generation}

\begin{figure*}[t]
\centerline{\includegraphics[width=\textwidth]{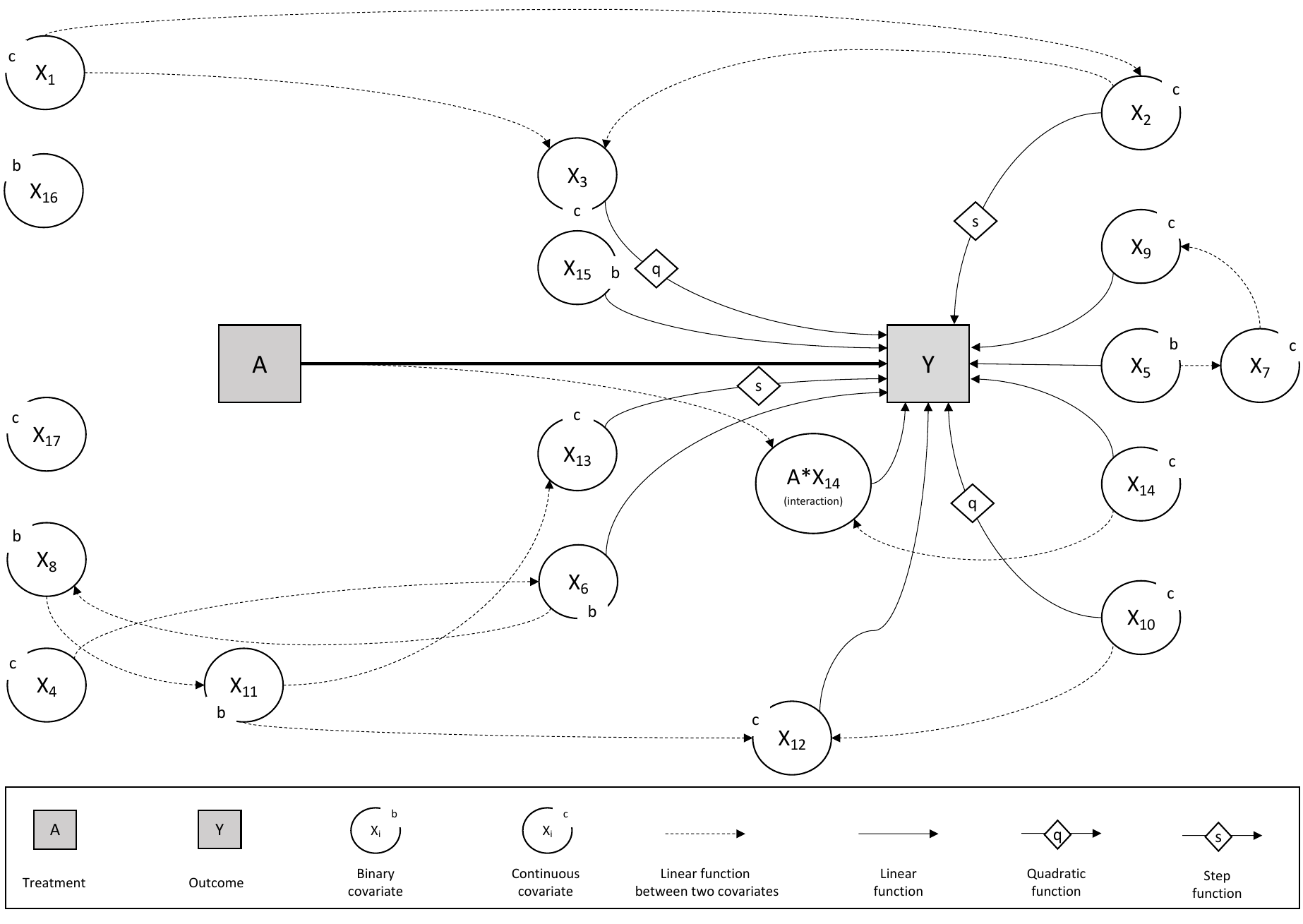}}
\caption{The causal relationships between the treatment arm, the covariates and the outcome for the complex scenario.\label{fig1}}
\end{figure*}

To evaluate the performances of the GC with ML,  we simulated two distinct scenarios inspired by our previous work.\cite{le_borgne_g-computation_2021} We first considered a complex scenario in which the estimation of the outcome model was not straightforward.  As illustrated in Figure~\ref{fig1},  it  involved 17 continuous or binary covariates with a complex structure of dependence.  Supplementary Table~\ref{tabS1}  presents the models used consecutively to generate the covariates.    Since the large majority of RCTs are designed to balance the allocation in each arm, often using a 1:1 randomization, $A$ was generated using Bernoulli distribution  with a probability equal to 0.5 and independently from $X=(X_1, ...,X_{17})$.   We then obtained the outcome $Y$ using Bernoulli distribution with a probability  equal to a logistic function of $A$ and $X$ (Supplementary Table~\ref{tabS1}).   In this outcome model,  one covariate interacted with the treatment,  two had step functions,  two followed quadratic functions,  and six were linear.   We considered three marginal treatment effects: $mOR = (1.0,  1.3, 1.9)$ and three sample sizes ($n = 60$,  $100$, $200$). For each of the 6 scenarios,  we generated 10,000 data sets. \cite{chatton_g-computation_2020}

\begin{figure*}[t]
\centerline{\includegraphics[width=0.65\textwidth]{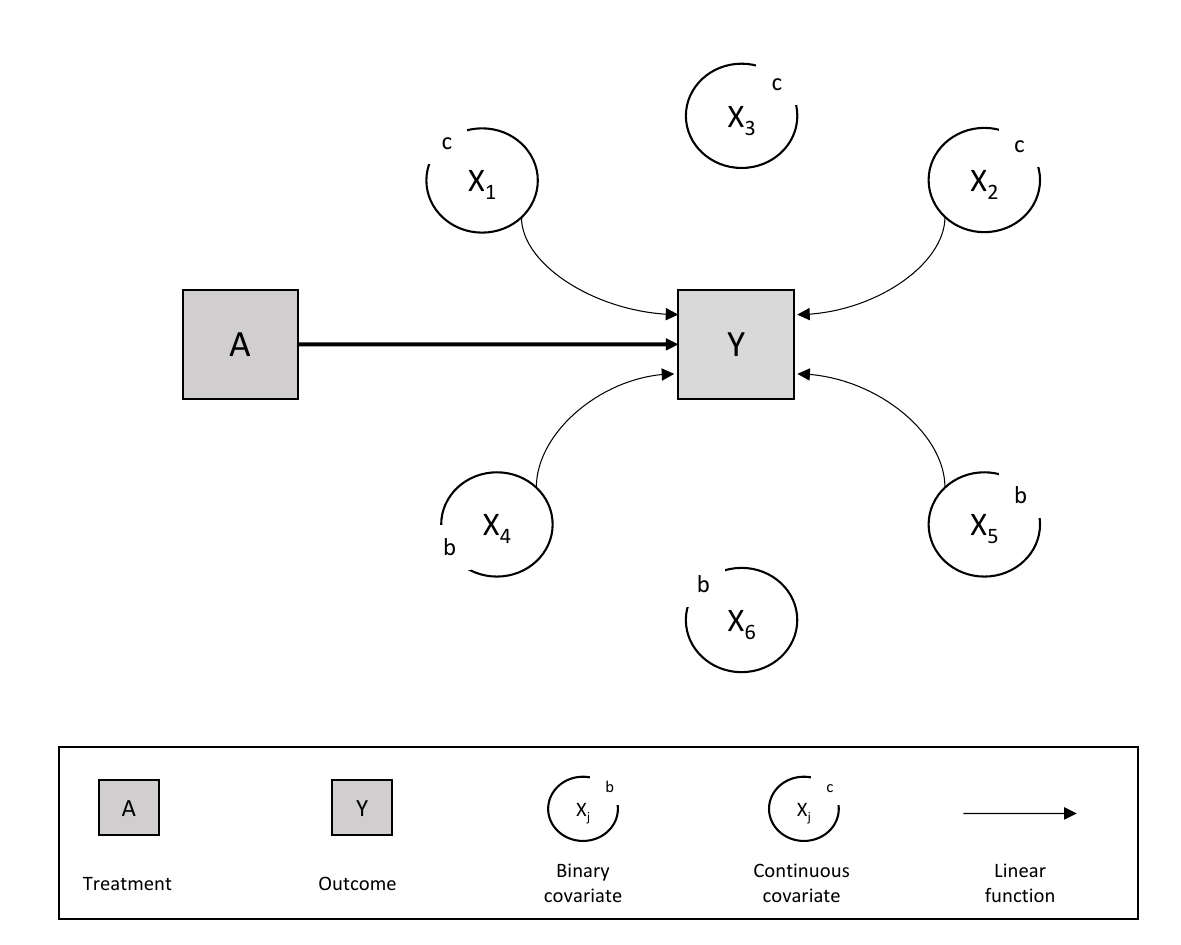}}
\caption{The causal relationships between the treatment arm, the covariates and the outcome for the simple  scenario.\label{fig2}}
\end{figure*}

We thereafter considered a simpler scenario,  i.e.  with a lower risk of model miss-specification.  As illustrated in Figure~\ref{fig2} and Supplementary Table~\ref{tabS2},    it  involved 3 binary and 3 continuous covariates.  No interaction was introduced and the log-linearity  assumption was respected. We considered three marginal treatment effects: $mOR = (1.0, 1.3, 1.9)$ with a lower proportion of individuals with the outcome $Y=1$ (estimated asymptotic values are provided in the footnotes of all tables). The other characteristics of the simulations were similar to the complex scenario. Furthermore, to extend the simulated results and closely match the application data, we considered the same simple scenario with a model for outcome prediction which had reduced predictive performances, i.e. smaller regression coefficients (Supplementary Table~\ref{tabS2}). Additionally, in this simpler scenario with reduced predictive performance, we evaluated the impact of excluding interactions and B-splines on the performance for the penalized logistic regressions.

\subsubsection{Applications}

In addition to the simulated scenarios, we applied our methods to two individual randomized trials. Due to missing data in the prognostic factors of the outcome, a complete case analysis was firstly considered. Secondly, multiple imputations were performed. Fifty imputed datasets were considered, with predictive mean matching method for continuous data and logistic regression for categorical variables. The MI BOOT method, as described by Schomaker and Heumann,\cite{schomaker_bootstrap_2018} was applied to combine multiple imputations with bootstrap resampling. All the baseline variables reported in the descriptive tables were considered.

\subsubsection{Performance criteria}

We evaluated several criteria defined in supplementary section ~\ref{app2}: mean bias (MB),  variance estimation bias (VEB),  root mean square error (RMSE),   empirical coverage rate of the nominal 95\% confidence interval,  error rates (type I when $mOR=1.0$ and type II otherwise) and reduction in sample  size (RSS).\cite{morris_using_2019}   The theoretical values of the marginal effects were estimated by the mean of the unadjusted estimations for 1,000,000 simulated data sets. For the applications, the area under the curve (AUC) was estimated. \\

\section{Results}

\subsection{Complex scenario}


As illustrated in Table~\ref{tab1},  because of the random allocation of the treatment, the absence of true confounders leads to MB values close to 0 for the unadjusted analyses.  Regardless the sample size or the treatment effect, the two penalized regressions performed better than the algorithm-based methods  (neural network, support vector machine,  and super learner).  More precisely,  the MB of the logarithm of the mOR ranged from 0.00 to 0.05 for Elasticnet,  and from 0.01 to 0.25 for Lasso logistic regression. The differences between the methods were even more important with regard to the risk difference.  The maximum MB of the penalized regressions was 0.61\%,  versus 8.18\% for the neural network,  6.70\% for the support vector machine,  and 4.51\% for the super learner.  The maximum MB values were obtained for the smaller sample size ($n=60$,  i.e. 30 individuals per arm) and the larger treatment effect ($mOR=1.9$).


The two penalized regressions best-respected the 5\% nominal type I error rate. The type II error rate significantly decreased compared to the unadjusted results.  For instance,  for $n=60$ and $mOR = 1.3$,  the type II error rate decreased from 92.31\%  to 89.78\% when using the Elasticnet and to 91.43\% for the Lasso logistic regressions.  It corresponded to a 28.7\% and 21.7\% RSS, respectively. When $n=200$ and $mOR = 1.9$, the type II error rate decreased from 42.58\%  to 26.07\% for the Elasticnet  logistic regression  and to 27.19\% for the Lasso logistic regression. It corresponded to a 31.1\% and 27.5\% RSS, respectively. For the two methods, the values of the RSS did not vary according to the effect size or sample size. In contrast, neural network showed a significant increase in the type II error rate for all effect and sample sizes. For example, with $n=200$ and $mOR = 1.9$, the type II error rate increased to 48.5\%, with a corresponding RSS of -30.9\%.

The rest of the results are regarding the mOR.  The Elasticnet logistic regression offered the best results in terms of variance bias, which was negligible for $n=200$,  but a small underestimation of the variance for $n=60$  ($VEB = -3.02\%$ for $mOR = 1.9$,  and $VEB = -2.54\%$ for $mOR = 1.3$).  The VEB was reasonable too for the Lasso logistic regression,  but a small overestimation for all effect sizes.  For the three algorithm-based methods,   the simulations revealed an underestimation of variance, except under the null hypothesis with the $mOR=1.0$ for $n=200$ for support vector machine and super learner.

The smaller values of RMSE were achieved by the using the neural network, support vector machine,  and super learner. This is a direct consequence of the variance underestimation.  When considering the two penalized regressions,  their RMSE were  close,  but with a small advantage for the Lasso logistic regression, attributed to its small underestimation of the variance.

The Lasso logistic regression resulted in a  coverage rate close to the 95\% nominal value. More precisely,  it ranged from 93.9\% ($n=60$ and $mOR = 1.9$) to 95.2\% ($n=200$ and $mOR = 1.9$).   For the Elasticnet regression,  the coverage rates were also nominal for $n=200$,  but sub-nominal for the smaller sample size: up to 92.7\% for $mOR = 1.9$.  Always regarding the underestimation of the variances, the coverage rates were largely lower for the algorithm-based methods: from 94.1\% to 80.0\% for the neural network,  from 94.8\% to 80.3\% for the support vector machine,  and from 95.4\% to  91.0\% for the super learner.

\subsection{Simple scenario}
The results of the simple scenario are depicted in Table~\ref{tab2}. In this subsection, we will focus on the main similarities and differences compared to the complex scenario.


The two penalized regressions always outperformed alternative algorithms. Specifically, the MB of the logarithm of mOR ranged from 0.00 to 0.05 for Elasticnet and from 0.01 to 0.25 for Lasso logistic regressions. For $n=60$, the three algorithm-based methods had higher values.


In contrast to the complex scenario, the Elasticnet logistic regression had smaller type II error rates compared to the Lasso logistic regression.
For $n=200$ and $n=100$, the penalized regressions resulted in RSS values ranging from 23.3\% to 33.2\%. For $n=60$, the gain was even larger, averaging 50\%  (49.4\% and 53.6\% for Elasticnet versus 45.6\% and 51.0\% for Lasso).

The VEB values were higher compared to the complex scenario, this difference being partly due to the reduced proportion of individuals with $Y=1$ in this simple scenario.
The Elasticnet logistic regression offered the best results in terms of variance bias. Lasso logistic regression did not differ much. For the two penalized regressions, RMSE were almost identical, but with a small advantage for the Lasso.

For all the methods, the coverage rates were close to the 95\% nominal value, ranging from 93.1\% for support vector machine ($n=60$ and $mOR = 1.9$) to 97.4\% for neural network ($n=200$ and $mOR = 1.9$). For the Elasticnet logistic regression, the coverage rate ranged from 93.8\% to 94.9\%. The Lasso logistic regression did not suffer from this slight undercoverage, values ranged from 94.6\% to 95.4\%.

\subsection{Reduced predictive performance of the outcome model}\label{s1s-bis}
In the simple scenario with reduced predictive performance, the AUC of the outcome model was 0.67 compared to 0.89 in the previous results. In contrast to the previous results, the RMSE did not decrease with the GC based on penalized regressions. The GC was associated with an overestimation of the treatment effect variance, with this bias being higher for smaller sample sizes and smaller mOR values. For instance, the VEB of the Elasticnet was 2.02\% for $mOR=1.5$ and $n=200$, versus 6.27\% for $n=100$ and 12.01\% for $n=60$ (Supplementary Table~\ref{tabs3}). However, after excluding both interactions and B-splines from the penalized logistic regressions, a substantial decrease in VEB was observed, with an absolute maximum for the Elasticnet of $2.27\%$ for $n=100$ and $mOR=1.0$ (Supplementary Table~\ref{tabs3b}). The RMSE and coverage rates remained the same, but the RSS values, which had previously been negative, became entirely positive and increased as the sample size decreased.

\section{Applications}

\subsection{Daclizumab versus antithymocyte globulin in kidney transplant recipients}

The objective of the TAXI trial was to elucidate whether there is any significant impact on the incidence of biopsy-proven acute rejection (principal outcome), delayed graft function, and graft loss at one year post-surgery between daclizumab (DAC) and antithymocyte globulin (ATG).  \cite{noel_daclizumab_2009}  This  1:1 RCT was  designed for a power of 80\%. Among the 227 enrolled patients, 113 were randomized in the ATG arm and 114 in the DAC arm.
Among the  prognostic factors (listed in Supplementary Table~\ref{tabS3}) one can observe several  differences between the two arms. For instance, the ATG arm consisted of 46.0\% males compared to 51.8\% in the DAC arm, leading to a standardized mean difference (SMD) of 11.5\%. The percentage of donor deaths from stroke was 83.7\% versus 89.5\%, respectively (SMD = 18.6\%). The cold ischemia time was on average 1.3 hours higher in the ATG arm (SMD = 19.5\%). While these factors were established as prognostic factors, they may constitute near-confounders. Six patients had at least one missing data on the selected prognostic factors, resulting in a complete case population for $n=221$ patients ($n=110$ with ATG and $n=111$ with DAC).

\begin{sidewaysfigure*}[t]
    \centering
    \includegraphics[width=\textheight , trim=5cm 5cm 5cm 1cm,clip]{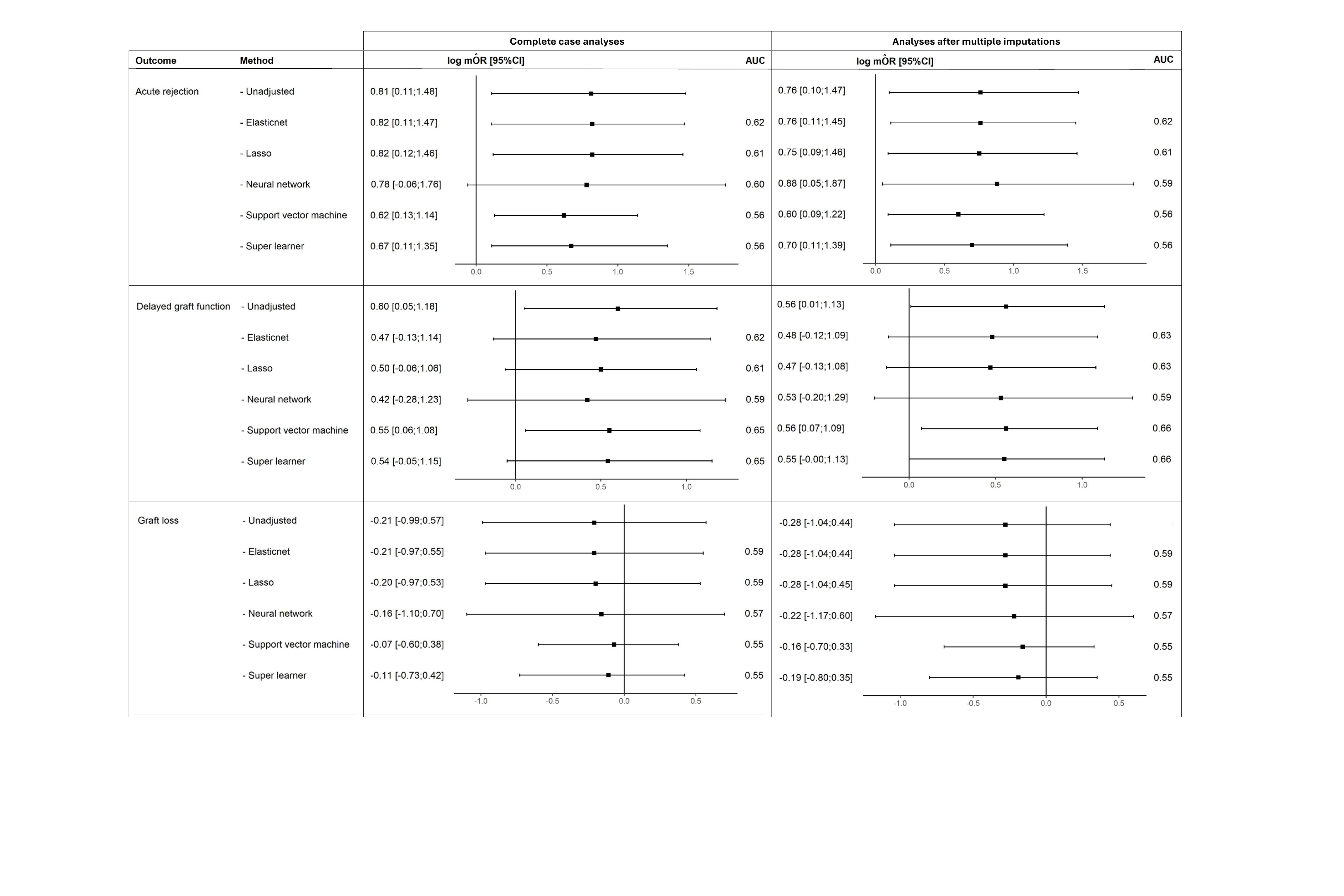}
    \caption{Results for the trial comparing daclizumab versus antithymocyte globulin in kidney transplant recipients with three different outcomes and two  populations: the complete case population ($n=221$) and the overall population ($n=227$, by using multiple imputation). AUC represents the mean of the area under the ROC curves of the outcome models}\label{sfig1}
\end{sidewaysfigure*}

As shown in Figure~\ref{sfig1}, analyzing the delayed graft function outcome, the adjusted effects of the different methods were consistently smaller than the unadjusted effects. For the other two outcomes, acute rejection and graft loss, the effects estimated using penalized regressions were closest to the unadjusted effects. In contrast, the adjusted results from the support vector machine and the super learner varied from the unadjusted effect (but were reported with a non-negligible mean bias in the simulation-based section). In terms of variance, we observed the smallest confidence intervals for both the support vector machine and the super learner. In contrast for the neural network, we observed the largest values. In agreement with the simulations, we observed a tiny reduction of the confidence intervals for the two penalized methods, this reduction being slightly higher for the Lasso regression.

\subsection{High-flow nasal oxygen versus standard oxygen face mask or noninvasive ventilation in patients with acute hypoxemic respiratory failure}

The FLORALI trial was an RCT with 1:1:1 randomization designed for a power of 80\%.  \cite{frat_high-flow_2015} The objectives were to compare the efficacy of high-flow oxygen versus standard oxygen therapy, and high-flow oxygen versus noninvasive ventilation in preventing endotracheal intubation within 28 days.
Among the 310 included patients,  106 were randomized to high-flow oxygen therapy, 94 patients to standard oxygen therapy, and 110 to noninvasive ventilation. The complete case analyses (i.e. no missing values among the prognostic factors listed in Supplementary Table~\ref{tabS5}) consisted of 105, 89 and 100 patients respectively. One can observe near-confounders, for example, the SAPS II was 2 points higher in the noninvasive ventilation versus the high-flow oxygen arms (SMD = 13.3\%). The highest SMD was observed for the PaO2 level between the standard and the high-flow oxygen arms with a difference of 7 mmHg.

\begin{sidewaysfigure*}[t]
    \centering
    \includegraphics[width=\textheight , trim=5cm 5cm 5cm 1cm,clip]{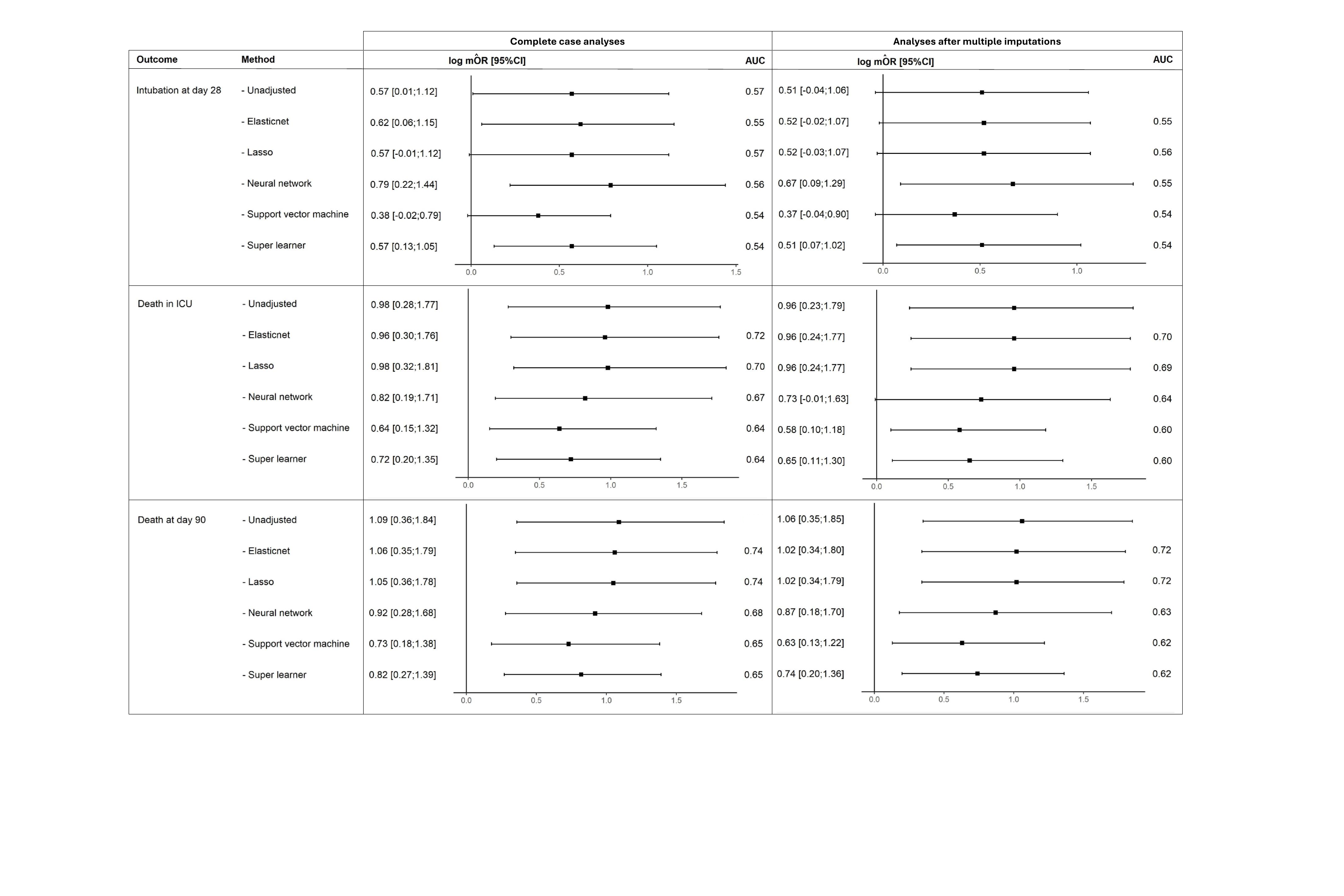}
    \caption{Results for the trial comparing high-flow nasal oxygen versus noninvasive ventilation in patients with acute hypoxemic respiratory failure with three different outcomes and two populations: the complete case population ($n=205$) and the overall population ($n=216$, by using multiple imputation). AUC represents the mean of the area under the ROC curves of the outcome models}\label{sfig3}
\end{sidewaysfigure*}

\begin{sidewaysfigure*}[t]
    \centering
    \includegraphics[width=\textheight , trim=5cm 5cm 5cm 1cm,clip]{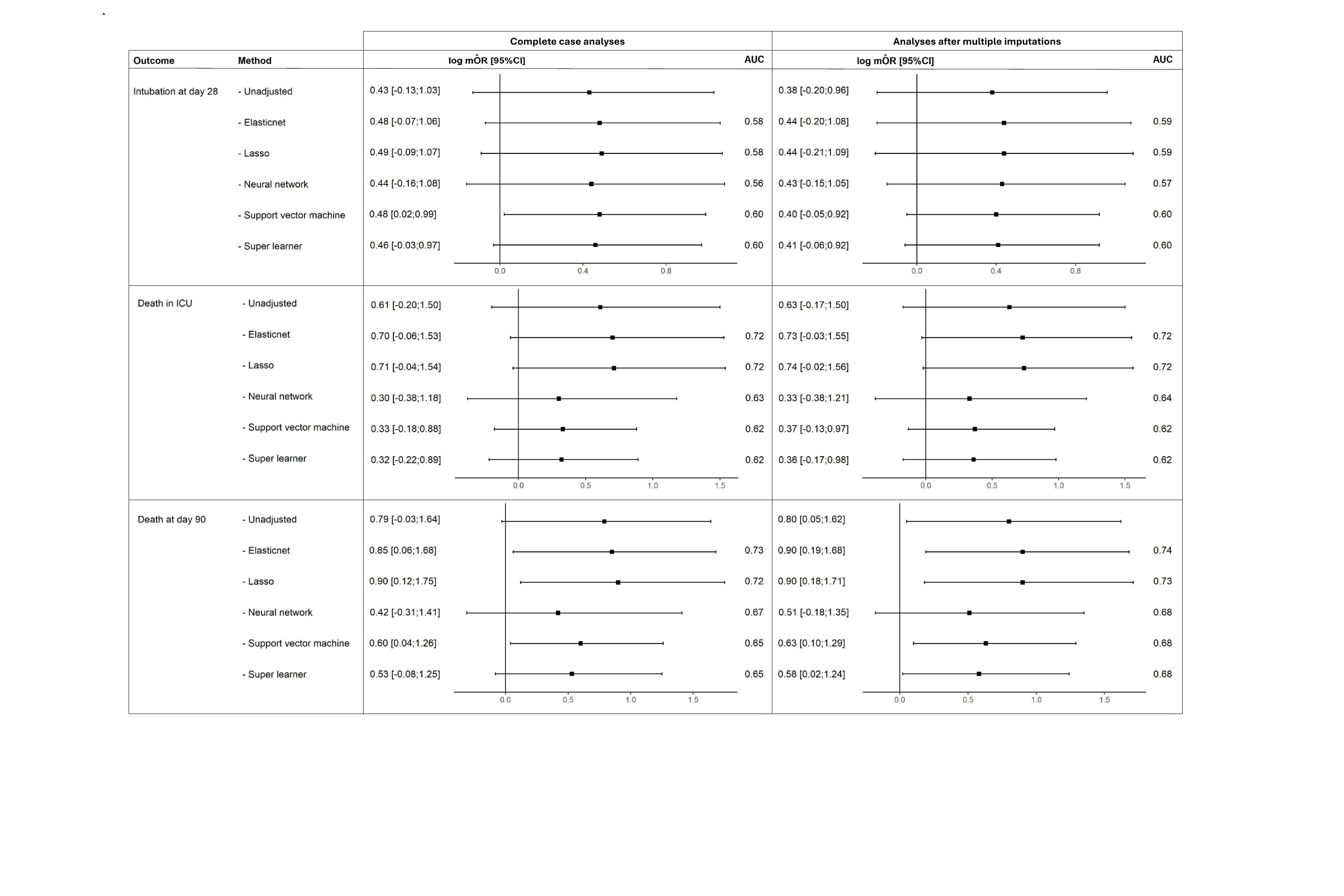}
    \caption{Results for the trial comparing high-flow nasal oxygen versus standard oxygen facemask in patients with acute hypoxemic respiratory failure with three different outcomes and two populations: the complete case population ($n=194$) and the overall population ($n=200$, by using multiple imputation). AUC represents the mean of the area under the ROC curves of the outcome models}\label{sfig2}
\end{sidewaysfigure*}

As for the previous application, the algorithm-based estimation differed from the unadjusted ones, while the estimated effects when using the penalized regressions were closer. We did not observe relevant differences between the complete case analyses and the multiple imputation analyses.
The main difference with the previous application is the increase in the AUC values, i.e. better predictive accuracy of the outcome models, especially for the mortality outcomes. The smallest confidence intervals when using the penalized methods were observed for AUC $>0.70$. For instance, when studying mortality in intensive care units, the log mOR was $0.73$ (95\% CI from $-0.03$ to $1.55$) for the Elasticnet regression, compared to $0.63$ (95\% CI from $-0.17$ to $1.50$) for the unadjusted analysis when using multiple imputations. The standard deviations were $0.40$ and $0.43$, respectively (Figure \ref{sfig3} and Supplementary Figure \ref{sfig2}).

\section{Discussion}

We started off by comparing methods to estimate the outcome model in GC in the context of RCTs with both complex and simple relationships between covariates and a binary outcome. The results from the simulations distinguished two categories of methods: penalized regressions (Lasso and Elasticnet) and the algorithm-based methods (neural network, support vector machine and super learner). Penalized regressions achieved a reduction in variance, although they introduced a slight increase in bias. The RSS ranged from a reduction equivalent to 17 patients (17.3\%, Lasso, $mOR=1.9$, $n=100$, complex scenario) to 32 patients (53.6\%, Elasticnet, $mOR=1.3$, $n=60$, simple scenario). These findings indicate that GC with penalized regressions is particularly effective in smaller trials with both complex and simple data structures.

The two applications also revealed the benefits of using GC with penalized outcome models. Indeed, we observed potential near-confounders, resulting in differences between the adjusted and unadjusted estimation of the marginal effects. As shown in the TAXI trial, prognostic variables, such as donor deaths from stroke, and cold ischemia time, were imbalanced. This may explain why the adjusted effects were all  smaller than the unadjusted ones when studying delayed graft function. We also described relevant reductions of the 95\% confidence intervals, especially when outcome models achieved significant predictive capacities. This suggests that even in RCTs with 100 patients per arm, adjusting for prognostic factors can significantly improve the estimation of treatment effects and reduce their confidence intervals. Even when baseline imbalances are minimal, a strong prognostic effect can lead to a substantial increase in power with covariate adjustment.\cite{steyerberg_clinical_2000,pocock_subgroup_2002}

As expected, as the size of the RCT increases, the patient allocation becomes more balanced, diminishing the effectiveness of the GC, shown to be more beneficial in smaller trials. Nevertheless, the two applications illustrated that near-confounders, were present even with more than 100 randomized patients per arm, while, systematic reviews, \cite{robinson_characteristics_2021, anthon_overall_2019} reported a majority of RCT with lower sample sizes.

Based on the simulations, the Elasticnet logistic regression offered the best results in terms of VEB, which was negligible for $n=200$ and slightly underestimated for $n=60$. The VEB was reasonable for the Lasso logistic regression, but small overestimations were observed. The mean bias was negligible for the two penalized approaches regardless of the sample size and the effect size. Note that, Tackney et al.\cite{tackney_comparison_2023} found that GC can underestimate the standard deviation in small sample sizes ($n=100$). In our context, we found that this underestimation occurs only for the algorithm-based machine learning methods, whereas penalized regression methods demonstrate accurate estimation of the standard deviation. 

When the prognostic factors did not provide sufficient predictive capacities for the outcome model, the adjusted and unadjusted results tended to be similar in terms of effect estimations. In this situation, we even reported an increase of the variance estimation for small sample sizes ($n\leq 100$). However, even though the RSS was negative, the RMSE was similar to the unadjusted for all simulated sample sizes with their respective RMSE ranging from 0.30 for $n=200$ to 0.60 for $n=60$. This means that the GC based on penalized methods reduced the error related to any near-confounders, but at the same time increases the estimation of effect variance. 
In contrast, the analyses that excluded interactions and B-splines for this scenario, revealed a notable reduction in VEB. Additionally, the penalized methods showed a positive RSS across all sample sizes. These results call for using simple outcome models for reduced sample sizes.

For the three algorithm-based methods, the simulations revealed variance underestimations. These results were corroborated with the applications, especially the RCT in patients with respiratory failure (Figure~\ref{sfig2}). One can also have serious concerns with regard to the effect of the estimations, being often far from those obtained with no adjustments.
The results of our simulations, in accordance with the findings of Le Borgne et al.\cite{le_borgne_g-computation_2021}, indicate that penalized methods perform better in simplistic scenarios, while algorithm-based methods are more appropriate in complex situations that necessitate larger sample sizes for training. \cite{van_der_ploeg_modern_2014}

In the applications, some of the covariates had missing data. In this article, our goal was not to study different imputation methods. However, the results illustrated that imputations added additional variability, while complete case analyses, reduced the sample, both resulting in a decrease in power.
It outlines the importance of accurately collecting prognostic factors to prevent loss of power, as this would contradict the initial goal to increase power.

While this study focused on the context of superiority trials, the methods explored could also be valuable in non-inferiority trials. In non-inferiority designs, the intention-to-treat analysis can introduce bias, making per protocol analysis more suitable. \cite{schumi_through_2011} Nicholas et al. demonstrated that adjustment methods can improve estimation accuracy in non-inferiority trials as well.\cite{nicholas_impact_2015} These findings suggest a broader applicability of GC with penalized methods across different trial designs.

Our study had some limitations. First, we only used GC and did not compare to other methods for marginal effect estimation, such as propensity scores or doubly robust estimators. These alternatives are based on the modelling of the treatment allocation, which can be counterintuitive in the context of RCTs. 
Secondly, we used a single outcome model for both arms. Although stratifying the estimations of outcome models for each arm could by design account for interactions between treatment and covariates,\cite{kunzel_metalearners_2019} this approach may increase variability, especially in small sample sizes, while we specifically aim to increase the power.

Thirdly, the results from our simulations may not be generalized to every possible scenario, even though we tried to explore a wide range with 27 scenarios. Fourthly, our study was limited to binary outcomes. Nonetheless, the applicability of GC can also be extended to time-to-event analyses. \cite{chatton_g-computation_2022}
Fifthly and lastly, we used the AUC as the metric to estimate the super learner which included support vector machine as one of the learners. AUC and support vector machine are particularly relevant in the context of classification. Moreover, the weights of the super learner indicated that support vector machine and neural network were higher compared to the two penalized methods (data not shown). The results of our work suggest that further developments are needed to evaluate if other super learners may be more adapted in this context of GC, i.e. outcome regression.

To conclude, the results show that the GC with penalized models, particularly Elasticnet with splines when appropriate, represents a relevant approach for increasing the power of RCTs and accounting for  near-confounders. This approach is particularly effective when the outcome model has significant predictive capacities, i.e., includes relevant prognostic factors associated with the outcome. 
Moreover, GC with penalized methods remains robust even with very small sample sizes ($n=60$). 
However, it is crucial to avoid complex outcome models when the sample size is limited.

\section*{Author contributions}
Y.F.  conceived the presented idea.  R.L.  and J.D.K. performed the simulations and analysed the data.  All authors discussed the results and contributed to the final manuscript.

\section*{Acknowledgments}
We acknowledge the principal investigators of the two trials, Jean-Pierre Frat, Arnaud Thille, Marc Hazzan and Christian Noël, whose data were utilized in this study. Their comprehensive data collection and diligent research efforts were crucial in the applications of our work. Computing time for this study was provided by the facilities of the MCIA (Mésocentre de Calcul Intensif Aquitain).

\section*{Financial disclosure}
This work was supported by a French government grant managed by the Agence Nationale de la Recherche under the France 2030 program, reference ANR-22-PESN-0003 SMATCH.

\section*{Conflict of interest}
The authors declare no potential conflict of interests.

\bibliographystyle{abbrv}
\bibliography{bib}

\begin{thebibliography}{10}

\bibitem{noauthor_consort_2010}
{CONSORT} 2010.
\newblock {\em Lancet (London, England)}, 375(9721):1136, Apr. 2010.

\bibitem{anthon_overall_2019}
C.~T. Anthon, A.~Granholm, A.~Perner, J.~H. Laake, and M.~H. Møller.
\newblock Overall bias and sample sizes were unchanged in {ICU} trials over
  time: a meta-epidemiological study.
\newblock {\em Journal of Clinical Epidemiology}, 113:189--199, Sept. 2019.

\bibitem{austin_statistical_2010}
P.~C. Austin.
\newblock Statistical {Criteria} for {Selecting} the {Optimal} {Number} of
  {Untreated} {Subjects} {Matched} to {Each} {Treated} {Subject} {When} {Using}
  {Many}-to-{One} {Matching} on the {Propensity} {Score}.
\newblock {\em American Journal of Epidemiology}, 172(9):1092--1097, Nov. 2010.

\bibitem{center_for_drug_evaluation_and_research_fda_adjusting_2023}
{Center for Drug Evaluation and Research, FDA}.
\newblock Adjusting for {Covariates} in {Randomized} {Clinical} {Trials} for
  {Drugs} and {Biological} {Products}, May 2023.

\bibitem{chatton_g-computation_2022}
A.~Chatton, F.~L. Borgne, C.~Leyrat, and Y.~Foucher.
\newblock G-computation and doubly robust standardisation for continuous-time
  data: {A} comparison with inverse probability weighting.
\newblock {\em Statistical Methods in Medical Research}, 31(4):706--718, Apr.
  2022.

\bibitem{chatton_g-computation_2020}
A.~Chatton, F.~Le~Borgne, C.~Leyrat, F.~Gillaizeau, C.~Rousseau, L.~Barbin,
  D.~Laplaud, M.~Léger, B.~Giraudeau, and Y.~Foucher.
\newblock G-computation, propensity score-based methods, and targeted maximum
  likelihood estimator for causal inference with different covariates sets: a
  comparative simulation study.
\newblock {\em Scientific Reports}, 10(1):9219, June 2020.

\bibitem{colson_optimizing_2016}
K.~E. Colson, K.~E. Rudolph, S.~C. Zimmerman, D.~E. Goin, E.~A. Stuart,
  M.~v.~d. Laan, and J.~Ahern.
\newblock Optimizing matching and analysis combinations for estimating causal
  effects.
\newblock {\em Scientific Reports}, 6(1):23222, Mar. 2016.

\bibitem{daniel_making_2021}
R.~Daniel, J.~Zhang, and D.~Farewell.
\newblock Making apples from oranges: {Comparing} noncollapsible effect
  estimators and their standard errors after adjustment for different covariate
  sets.
\newblock {\em Biometrical Journal}, 63(3):528--557, 2021.

\bibitem{european_medical_agency_guideline_2015}
{European Medical Agency}.
\newblock {\em Guideline on adjustment for baseline covariates in clinical
  trials}.
\newblock 2015.
\newblock Report No.: EMA/CHMP/295050/2013.

\bibitem{frat_high-flow_2015}
J.-P. Frat, A.~W. Thille, A.~Mercat, C.~Girault, S.~Ragot, S.~Perbet, G.~Prat,
  T.~Boulain, E.~Morawiec, A.~Cottereau, J.~Devaquet, S.~Nseir, K.~Razazi,
  J.-P. Mira, L.~Argaud, J.-C. Chakarian, J.-D. Ricard, X.~Wittebole,
  S.~Chevalier, A.~Herbland, M.~Fartoukh, J.-M. Constantin, J.-M. Tonnelier,
  M.~Pierrot, A.~Mathonnet, G.~Béduneau, C.~Delétage-Métreau, J.-C.~M.
  Richard, L.~Brochard, R.~Robert, {FLORALI Study Group}, and {REVA Network}.
\newblock High-flow oxygen through nasal cannula in acute hypoxemic respiratory
  failure.
\newblock {\em The New England Journal of Medicine}, 372(23):2185--2196, June
  2015.

\bibitem{funk_doubly_2011}
M.~J. Funk, D.~Westreich, C.~Wiesen, T.~Stürmer, M.~A. Brookhart, and
  M.~Davidian.
\newblock Doubly {Robust} {Estimation} of {Causal} {Effects}.
\newblock {\em American Journal of Epidemiology}, 173(7):761--767, Apr. 2011.

\bibitem{greenland_interpretation_1987}
S.~Greenland.
\newblock Interpretation and choice of effect measures in epidemiologic
  analyses.
\newblock {\em American Journal of Epidemiology}, 125(5):761--768, May 1987.

\bibitem{kunzel_metalearners_2019}
S.~R. Künzel, J.~S. Sekhon, P.~J. Bickel, and B.~Yu.
\newblock Metalearners for estimating heterogeneous treatment effects using
  machine learning.
\newblock {\em Proceedings of the National Academy of Sciences of the United
  States of America}, 116(10):4156--4165, Mar. 2019.

\bibitem{le_borgne_g-computation_2021}
F.~Le~Borgne, A.~Chatton, M.~Léger, R.~Lenain, and Y.~Foucher.
\newblock G-computation and machine learning for estimating the causal effects
  of binary exposure statuses on binary outcomes.
\newblock {\em Scientific Reports}, 11(1):1435, Jan. 2021.

\bibitem{lendle_targeted_2013}
S.~D. Lendle, B.~Fireman, and M.~J. {van der Laan}.
\newblock Targeted maximum likelihood estimation in safety analysis.
\newblock {\em Journal of Clinical Epidemiology}, 66(8, Supplement):S91--S98,
  Aug. 2013.

\bibitem{lingsma_covariate_2010}
H.~Lingsma, B.~Roozenbeek, and E.~Steyerberg.
\newblock Covariate adjustment increases statistical power in randomized
  controlled trials.
\newblock {\em Journal of Clinical Epidemiology}, 63(12):1391, Dec. 2010.

\bibitem{leger_causal_2022}
M.~Léger, A.~Chatton, F.~Le~Borgne, R.~Pirracchio, S.~Lasocki, and Y.~Foucher.
\newblock Causal inference in case of near-violation of positivity: comparison
  of methods.
\newblock {\em Biometrical Journal. Biometrische Zeitschrift},
  64(8):1389--1403, Dec. 2022.

\bibitem{morris_using_2019}
T.~P. Morris, I.~R. White, and M.~J. Crowther.
\newblock Using simulation studies to evaluate statistical methods.
\newblock {\em Statistics in Medicine}, 38(11):2074--2102, May 2019.

\bibitem{nicholas_impact_2015}
K.~Nicholas, S.~D. Yeatts, W.~Zhao, J.~Ciolino, K.~Borg, and V.~Durkalski.
\newblock The impact of covariate adjustment at randomization and analysis for
  binary outcomes: understanding differences between superiority and
  noninferiority trials.
\newblock {\em Statistics in Medicine}, 34(11):1834--1840, May 2015.

\bibitem{noel_daclizumab_2009}
C.~Noël, D.~Abramowicz, D.~Durand, G.~Mourad, P.~Lang, M.~Kessler,
  B.~Charpentier, G.~Touchard, F.~Berthoux, P.~Merville, N.~Ouali, J.-P.
  Squifflet, F.~Bayle, K.~M. Wissing, and M.~Hazzan.
\newblock Daclizumab versus antithymocyte globulin in high-immunological-risk
  renal transplant recipients.
\newblock {\em Journal of the American Society of Nephrology: JASN},
  20(6):1385--1392, June 2009.

\bibitem{pocock_subgroup_2002}
S.~J. Pocock, S.~E. Assmann, L.~E. Enos, and L.~E. Kasten.
\newblock Subgroup analysis, covariate adjustment and baseline comparisons in
  clinical trial reporting: current practice and problems.
\newblock {\em Statistics in Medicine}, 21(19):2917--2930, Oct. 2002.

\bibitem{riley_penalization_2021}
R.~D. Riley, K.~I.~E. Snell, G.~P. Martin, R.~Whittle, L.~Archer, M.~Sperrin,
  and G.~S. Collins.
\newblock Penalization and shrinkage methods produced unreliable clinical
  prediction models especially when sample size was small.
\newblock {\em Journal of Clinical Epidemiology}, 132:88--96, Apr. 2021.

\bibitem{robinson_characteristics_2021}
N.~B. Robinson, S.~Fremes, I.~Hameed, M.~Rahouma, V.~Weidenmann, M.~Demetres,
  M.~Morsi, G.~Soletti, A.~Di~Franco, M.~A. Zenati, S.~G. Raja, D.~Moher,
  F.~Bakaeen, J.~Chikwe, D.~L. Bhatt, P.~Kurlansky, L.~N. Girardi, and
  M.~Gaudino.
\newblock Characteristics of {Randomized} {Clinical} {Trials} in {Surgery}
  {From} 2008 to 2020: {A} {Systematic} {Review}.
\newblock {\em JAMA network open}, 4(6):e2114494, June 2021.

\bibitem{rubin_estimating_1974}
D.~B. Rubin.
\newblock Estimating causal effects of treatments in randomized and
  nonrandomized studies.
\newblock {\em Journal of Educational Psychology}, 66(5):688--701, 1974.
\newblock Place: US Publisher: American Psychological Association.

\bibitem{schomaker_bootstrap_2018}
M.~Schomaker and C.~Heumann.
\newblock Bootstrap {Inference} {When} {Using} {Multiple} {Imputation}.
\newblock {\em Statistics in medicine}, 37(14):2252--2266, June 2018.

\bibitem{schumi_through_2011}
J.~Schumi and J.~T. Wittes.
\newblock Through the looking glass: understanding non-inferiority.
\newblock {\em Trials}, 12:106, May 2011.

\bibitem{senn_statistical_2021}
S.~Senn.
\newblock {\em Statistical issues in drug development}.
\newblock Statistics in practice. Wiley Blackwell, Hoboken, NJ, third edition
  edition, 2021.

\bibitem{snowden_implementation_2011}
J.~M. Snowden, S.~Rose, and K.~M. Mortimer.
\newblock Implementation of {G}-computation on a simulated data set:
  demonstration of a causal inference technique.
\newblock {\em American Journal of Epidemiology}, 173(7):731--738, Apr. 2011.

\bibitem{steyerberg_clinical_2000}
E.~W. Steyerberg, P.~M. Bossuyt, and K.~L. Lee.
\newblock Clinical trials in acute myocardial infarction: should we adjust for
  baseline characteristics?
\newblock {\em American Heart Journal}, 139(5):745--751, May 2000.

\bibitem{tackney_comparison_2023}
M.~S. Tackney, T.~Morris, I.~White, C.~Leyrat, K.~Diaz-Ordaz, and
  E.~Williamson.
\newblock A comparison of covariate adjustment approaches under model
  misspecification in individually randomized trials.
\newblock {\em Trials}, 24(1):14, Jan. 2023.

\bibitem{van_der_ploeg_modern_2014}
T.~{van der Ploeg}, P.~C. Austin, and E.~W. Steyerberg.
\newblock Modern modelling techniques are data hungry: a simulation study for
  predicting dichotomous endpoints.
\newblock {\em BMC medical research methodology}, 14:137, Dec. 2014.

\bibitem{williamson_variance_2014}
E.~J. Williamson, A.~Forbes, and I.~R. White.
\newblock Variance reduction in randomised trials by inverse probability
  weighting using the propensity score.
\newblock {\em Statistics in Medicine}, 33(5):721--737, Feb. 2014.

\bibitem{zeng_propensity_2021}
S.~Zeng, F.~Li, R.~Wang, and F.~Li.
\newblock Propensity score weighting for covariate adjustment in randomized
  clinical trials.
\newblock {\em Statistics in Medicine}, 40(4):842--858, 2021.

\end{thebibliography}

\section*{Appendix}

\subsection*{Supplementary data}

\newpage

\begin{center}
\begin{table*}[!ht]%
\scriptsize
\caption{Results of the simulations for the complex scenario.\label{tab1}}
\begin{tabular*}{\textwidth}{@{\extracolsep\fill}c|c|c|rrrr|rrrrr}
\toprule
\multirow{2}{*}{\makecell{\textbf{Sample} \\ \textbf{size}}}  &  \multirow{2}{*}{\textbf{mOR}\tnote{$^*$} } & \multirow{2}{*}{\textbf{Method} } & \multicolumn{4}{c|}{\textbf{Mean bias}} & \multicolumn{5}{c}{\boldmath$\log\hat{mOR}$} \\\cline{4-7}\cline{8-12}
         &  &  & \boldmath$\hat{\pi}_0$ & \boldmath$\hat{\pi}_1$ &  \boldmath$\log\hat{mOR}$ &\boldmath$\hat{\Delta}$ & \textbf{RMSE} & \textbf{VEB} & \textbf{Coverage} & \textbf{Error}\tnote{$^\dagger$} & \textbf{RSS } \\ \hline
        \multirow{18}{*}{$n= 200$}& \multirow{6}{*}{1.9} & Unadjusted & 0.03\% & -0.02\% & 0.0044 & -0.05\% & 0.29 & -0.43\% & 94.57\% & 42.58\% & 0.00\% \\
        ~ & ~ & Elasticnet & -0.10\% & 0.03\% & 0.0108 & 0.13\% & 0.24 & 0.61\% & 94.54\% & 26.07\% & 31.07\% \\
        ~ & ~ & Lasso & -0.11\% & -0.07\% & 0.0071 & 0.04\% & 0.24 & 3.59\% & 95.24\% & 27.19\% & 27.53\% \\
        ~ & ~ & Neural network & 1.36\% & -2.27\% & -0.1439 & -3.63\% & 0.27 & -5.14\% & 93.07\% & 48.56\% & -30.94\% \\
        ~ & ~ & Support vector machine & 0.80\% & -1.78\% & -0.1034 & -2.59\% & 0.24 & -9.94\% & 90.88\% & 31.22\% & 18.18\% \\
        ~ & ~ & Super learner & 1.71\% & -0.58\% & -0.0890 & -2.29\% & 0.24 & -7.09\% & 92.22\% & 29.16\% & 22.86\% \\ \cline{2-12}

        ~ & \multirow{6}{*}{1.3} & Unadjusted & 0.00\% & -0.06\% & 0.0000 & -0.06\% & 0.29 & -1.19\% & 91.96\% & 86.69\% & 0.00\% \\
        ~ & ~ & Elasticnet & -0.11\% & -0.08\% & -0.0030 & 0.02\% & 0.24 & -0.42\% & 95.04\% & 82.20\% & 32.90\% \\
        ~ & ~ & Lasso & -0.13\% & -0.12\% & 0.0025 & 0.01\% & 0.24 & 2.23\% & 94.65\% & 83.34\% & 30.28\% \\
        ~ & ~ & Neural network & 0.55\% & -1.73\% & -0.0899 & -2.27\% & 0.23 & 3.16\% & 94.14\% & 92.03\% & -79.15\% \\
        ~ & ~ & Support vector machine & 0.73\% & -0.33\% & -0.0415 & -1.06\% & 0.21 & -3.04\% & 94.80\% & 83.42\% & 25.90\% \\
        ~ & ~ & Super learner & 1.37\% & 0.20\% & -0.0456 & -1.17\% & 0.21 & -3.01\% & 93.57\% & 83.94\% & 20.42\% \\ \cline{2-12}
        
        ~ & \multirow{6}{*}{1.0} & Unadjusted & -0.03\% & 0.02\% & 0.0020 & 0.05\% & 0.29 & -1.38\% & 94.09\% & 5.93\% & \multicolumn{1}{c}{-} \\ 
        ~ & ~ & Elasticnet & -0.13\% & -0.08\% & 0.0020 & 0.05\% & 0.24 & 0.21\% & 94.39\% & 5.61\% & \multicolumn{1}{c}{-} \\ 
        ~ & ~ & Lasso & -0.15\% & -0.09\% & 0.0024 & 0.06\% & 0.24 & 2.72\% & 94.83\% & 5.17\% & \multicolumn{1}{c}{-} \\ 
        ~ & ~ & Neural network & -0.05\% & -1.25\% & -0.0493 & -1.20\% & 0.21 & 10.24\% & 95.76\% & 4.24\% & \multicolumn{1}{c}{-} \\ 
        ~ & ~ & Support vector machine & 0.58\% & 0.67\% & 0.0036 & 0.09\% & 0.20 & -0.02\% & 94.20\% & 5.80\% & \multicolumn{1}{c}{-} \\ 
        ~ & ~ & Super learner & 1.10\% & 0.80\% & -0.0131 & -0.30\% & 0.20 & -0.05\% & 94.01\% & 5.99\% & \multicolumn{1}{c}{-} \\      \hline

         \multirow{18}{*}{$n = 100$} & \multirow{6}{*}{1.9} & Unadjusted & -0.12\% & 0.12\% & 0.0252 & 0.24\% & 0.43 & -0.78\% & 94.36\% & 65.52\% & 0.00\% \\
        ~ & ~ & Elasticnet & -0.47\% & 0.14\% & 0.0402 & 0.61\% & 0.39 & -1.20\% & 93.78\% & 55.35\% & 24.20\% \\
        ~ & ~ & Lasso & -0.54\% & -0.16\% & 0.0314 & 0.39\% & 0.38 & 3.40\% & 95.04\% & 58.35\% & 17.30\% \\ 
        ~ & ~ & Neural network & 2.27\% & -3.55\% & -0.2329 & -5.82\% & 0.36 & -14.83\% & 88.35\% & 72.49\% & -42.67\% \\
        ~ & ~ & Support vector machine & 1.47\% & -3.02\% & -0.1807 & -4.49\% & 0.33 & -19.20\% & 86.56\% & 57.08\% & 17.15\% \\
        ~ & ~ & Super learner & 2.36\% & -1.00\% & -0.1309 & -3.36\% & 0.32 & -7.56\% & 91.30\% & 58.35\% & 15.66\% \\  \cline{2-12}
        
        ~ & \multirow{6}{*}{1.3} & Unadjusted & -0.05\% & 0.03\% & 0.0087 & 0.08\% & 0.43 & -1.62\% & 94.31\% & 89.60\% & 0.00\% \\ 
        ~ & ~ & Elasticnet & -0.34\% & -0.16\% & 0.0127 & 0.18\% & 0.38 & -1.83\% & 93.69\% & 87.42\% & 25.21\% \\ 
        ~ & ~ & Lasso & -0.43\% & -0.29\% & 0.0115 & 0.14\% & 0.37 & 2.33\% & 94.66\% & 88.86\% & 20.08\% \\ 
        ~ & ~ & Neural network & 0.88\% & -2.30\% & -0.1277 & -3.19\% & 0.29 & -1.23\% & 92.89\% & 94.01\% & -135.62\% \\ 
        ~ & ~ & Support vector machine & 1.32\% & -0.48\% & -0.0713 & -1.80\% & 0.27 & -7.98\% & 91.57\% & 86.96\% & 22.21\% \\ 
        ~ & ~ & Super learner & 2.02\% & 0.44\% & -0.0623 & -1.57\% & 0.29 & -2.27\% & 93.37\% & 88.86\% & 10.02\% \\ \cline{2-12}
        
        ~ & \multirow{6}{*}{1.0} & Unadjusted & -0.00\% & 0.05\% & 0.0023 & 0.05\% & 0.42 & -0.21\% & 94.58\% & 5.43\% & \multicolumn{1}{c}{-} \\ 
        ~ & ~ & Elasticnet & -0.28\% & -0.19\% & 0.0034 & 0.08\% & 0.37 & -0.22\% & 93.98\% & 6.00\% & \multicolumn{1}{c}{-} \\ 
        ~ & ~ & Lasso & -0.35\% & -0.24\% & 0.0045 & 0.11\% & 0.36 & 3.99\% & 94.83\% & 5.16\% & \multicolumn{1}{c}{-} \\ 
        ~ & ~ & Neural network & -0.03\% & -1.32\% & -0.0544 & -1.28\% & 0.25 & 9.31\% & 95.52\% & 4.51\% & \multicolumn{1}{c}{-} \\ 
        ~ & ~ & Support vector machine & 1.09\% & 1.19\% & 0.0040 & 0.10\% & 0.25 & -3.94\% & 93.23\% & 6.78\% & \multicolumn{1}{c}{-} \\ 
        ~ & ~ & Super learner & 1.66\% & 1.46\% & -0.0118 & -0.20\% & 0.27 & 1.38\% & 94.30\% & 5.70\% & \multicolumn{1}{c}{-} \\   \hline

         \multirow{18}{*}{$n = 60$} & \multirow{6}{*}{1.9} & Unadjusted & 0.14\% & -0.04\% & 0.0261 & -0.18\% & 0.60 & 3.60\% & 93.99\% & 77.65\% & 0.00\% \\
        ~ & ~ & Elasticnet & -0.43\% & 0.05\% & 0.0523 & 0.48\% & 0.55 & -3.02\% & 92.72\% & 70.31\% & 29.69\% \\
        ~ & ~ & Lasso & -0.60\% & -0.41\% & 0.0405 & 0.19\% & 0.54 & 2.26\% & 93.94\% & 73.39\% & 22.23\% \\ 
        ~ & ~ & Neural network & 3.37\% & -4.81\% & -0.3349 & -8.18\% & 0.44 & -27.77\% & 80.05\% & 80.81\% & -27.63\% \\
        ~ & ~ & Support vector machine & 2.84\% & -3.86\% & -0.2759 & -6.70\% & 0.40 & -29.48\% & 80.30\% & 70.94\% & 27.33\% \\
        ~ & ~ & Super learner & 2.94\% & -1.57\% & -0.1831 & -4.51\% & 0.41 & -4.85\% & 90.99\% & 79.42\% & 15.36\% \\ \cline{2-12}
        
        ~ & \multirow{6}{*}{1.3} & Unadjusted & 0.07\% & 0.13\% & 0.0141 & 0.06\% & 0.57 & 1.23\% & 94.31\% & 92.31\% & 0.00\% \\
        ~ & ~ & Elasticnet & -0.40\% & -0.02\% & 0.0272 & 0.37\% & 0.53 & -2.54\% & 93.02\% & 89.78\% & 28.70\% \\
        ~ & ~ & Lasso & -0.57\% & -0.29\% & 0.0238 & 0.28\% & 0.52 & 2.64\% & 94.57\% & 91.43\% & 21.70\% \\
        ~ & ~ & Neural network & 1.39\% & -2.42\% & -0.1558 & -3.81\% & 0.30 & -2.79\% & 90.53\% & 94.56\% & -142.96\% \\
        ~ & ~ & Support vector machine & 1.80\% & -0.70\% & -0.1008 & -2.49\% & 0.30 & -8.42\% & 90.94\% & 89.40\% & 25.84\% \\
        ~ & ~ & Super learner & 2.18\% & 0.40\% & -0.0723 & -1.78\% & 0.35 & 6.75\% & 95.37\% & 94.24\% & 3.95\% \\  \cline{2-12}
        
        ~ & \multirow{6}{*}{1.0} & Unadjusted & -0.03\% & 0.06\% & 0.0041 & 0.08\% & 0.57 & 1.89\% & 94.05\% & 5.96\% & \multicolumn{1}{c}{-} \\ 
        ~ & ~ & Elasticnet & -0.44\% & -0.27\% & 0.0079 & 0.18\% & 0.53 & -2.80\% & 92.96\% & 7.03\% & \multicolumn{1}{c}{-} \\ 
        ~ & ~ & Lasso & -0.61\% & -0.41\% & 0.0087 & 0.19\% & 0.52 & 1.95\% & 94.09\% & 5.91\% & \multicolumn{1}{c}{-} \\ 
        ~ & ~ & Neural network & 0.04\% & -1.10\% & -0.0498 & -1.14\% & 0.26 & 9.73\% & 94.72\% & 5.28\% & \multicolumn{1}{c}{-} \\ 
        ~ & ~ & Support vector machine & 1.06\% & 1.13\% & 0.0033 & 0.07\% & 0.28 & -4.28\% & 92.62\% & 7.37\% & \multicolumn{1}{c}{-} \\ 
        ~ & ~ & Super learner & 1.57\% & 1.40\% & -0.0078 & -0.17\% & 0.34 & 7.91\% & 95.99\% & 4.01\% & \multicolumn{1}{c}{-} \\  
\bottomrule
\end{tabular*}
\begin{tablenotes}
\item[$^{\rm *}$] For $mOR=1.9$, the estimated theoretical values were $\pi_0 = 0.4604$,  $\pi_1 = 0.6106$,  $\log mOR = 0.6153$,  and $\pi_1$ - $\pi_0 = 0.1503$.  For $mOR=1.3$,  the estimated theoretical values were $\pi_0 = 0.4604$,  $\pi_1 = 0.5177$,  $\log mOR = 0.2322$,  and $\pi_1$ - $\pi_0 = 0.0574$. For $mOR=1.0$, the estimated theoretical values were $\pi_0 = 0.4603$,  $\pi_1 = 0.4603$, $\log mOR = 5.8\times 10^{-5}$,  and  $\pi_1$ - $\pi_0 = 1.5\times 10^{-5}$. Theoretical AUC of the outcome model in the simulations: 0.8904.
\item[$^{\rm \dagger}$]Type II error (i.e., 100 - power) for $mOR>1.0$, and type I error for $mOR=1.0$.\\
Abbreviations:  $mOR$, marginal odds ratio; RMSE, root mean square error; VEB, variance estimation bias; RSS,  reduction in sample size.
\end{tablenotes}
\end{table*}
\end{center}

\begin{center}
\begin{table*}[!ht]%
\scriptsize
\caption{Results of the simulations for the simple scenario.\label{tab2}}
\begin{tabular*}{\textwidth}{@{\extracolsep\fill}c|c|c|rrrr|rrrrr}
\toprule
\multirow{2}{*}{\makecell{\textbf{Sample} \\ \textbf{size}}}  &  \multirow{2}{*}{\textbf{mOR}\tnote{$^*$} } & \multirow{2}{*}{\textbf{Method} } & \multicolumn{4}{c|}{\textbf{Mean bias}} & \multicolumn{5}{c}{\boldmath$\log\hat{mOR}$} \\\cline{4-7}\cline{8-12}
         &  &  & \boldmath$\hat{\pi}_0$ & \boldmath$\hat{\pi}_1$ &  \boldmath$\log\hat{mOR}$ &\boldmath$\hat{\Delta}$ & \textbf{RMSE} & \textbf{VEB} & \textbf{Coverage} & \textbf{Error}\tnote{$^\dagger$} & \textbf{RSS } \\ \hline
        \multirow{18}{*}{$n= 200$}& \multirow{6}{*}{1.9} & Unadjusted & 0.05\% & 0.02\% & 0.0081 & -0.03\% & 0.32 & -0.24\% & 94.40\% & 43.42\% & 0.00\% \\ 
        ~ & ~ & Elasticnet & -0.10\% & 0.09\% & 0.0171 & 0.19\% & 0.27 & 1.62\% & 94.74\% & 26.46\% & 30.68\% \\ 
        ~ & ~ & Lasso & -0.06\% & 0.08\% & 0.0149 & 0.14\% & 0.27 & 3.99\% & 95.27\% & 27.59\% & 27.97\% \\ 
        ~ & ~ & Neural network & 0.16\% & -1.00\% & -0.0387 & -1.16\% & 0.28 & 20.09\% & 97.43\% & 48.95\% & -23.53\% \\ 
        ~ & ~ & Support vector machine & 2.35\% & 1.72\% & -0.0451 & -0.64\% & 0.27 & 5.01\% & 95.61\% & 37.12\% & 9.74\% \\ 
        ~ & ~ & Super learner & 1.37\% & 0.80\% & -0.0311 & -0.57\% & 0.26 & 4.69\% & 95.38\% & 30.69\% & 22.10\% \\  \cline{2-12}
        
        ~ & \multirow{6}{*}{1.3} & Unadjusted & 0.02\% & 0.05\% & 0.0057 & 0.02\% & 0.34 & -0.52\% & 94.39\% & 87.44\% & 0.00\% \\ 
        ~ & ~ & Elasticnet & -0.10\% & 0.10\% & 0.0138 & 0.20\% & 0.28 & 1.87\% & 94.92\% & 84.09\% & 32.45\% \\ 
        ~ & ~ & Lasso & -0.06\% & 0.12\% & 0.0128 & 0.18\% & 0.28 & 3.93\% & 95.23\% & 84.99\% & 29.96\% \\ 
        ~ & ~ & Neural network & -0.13\% & -0.79\% & -0.0268 & -0.66\% & 0.28 & 19.27\% & 97.18\% & 92.93\% & -35.07\% \\ 
        ~ & ~ & Support vector machine & 1.99\% & 1.74\% & -0.0197 & -0.24\% & 0.27 & 7.76\% & 96.17\% & 88.79\% & 9.19\% \\ 
        ~ & ~ & Super learner & 1.20\% & 1.01\% & -0.0121 & -0.18\% & 0.26 & 5.42\% & 95.48\% & 86.54\% & 21.85\% \\  \cline{2-12}
        
        ~ & \multirow{6}{*}{1.0} & Unadjusted & 0.01\% & 0.08\% & 0.0034 & 0.07\% & 0.34 & 0.09\% & 94.42\% & 5.56\% & \multicolumn{1}{c}{-} \\ 
        ~ & ~ & Elasticnet & -0.11\% & 0.12\% & 0.0118 & 0.22\% & 0.29 & 3.04\% & 94.79\% & 5.21\% & \multicolumn{1}{c}{-} \\ 
        ~ & ~ & Lasso & -0.06\% & 0.14\% & 0.0106 & 0.20\% & 0.28 & 5.47\% & 95.27\% & 4.73\% & \multicolumn{1}{c}{-} \\ 
        ~ & ~ & Neural network & -0.29\% & -0.63\% & -0.0207 & -0.34\% & 0.29 & 20.78\% & 97.33\% & 2.68\% & \multicolumn{1}{c}{-} \\ 
        ~ & ~ & Support vector machine & 1.64\% & 1.72\% & 0.0042 & 0.08\% & 0.27 & 9.58\% & 96.04\% & 3.96\% & \multicolumn{1}{c}{-} \\ 
        ~ & ~ & Super learner & 1.05\% & 1.12\% & 0.0023 & 0.07\% & 0.26 & 6.90\% & 95.41\% & 4.60\% & \multicolumn{1}{c}{-} \\       \hline

         \multirow{18}{*}{$n = 100$} & \multirow{6}{*}{1.9} & Unadjusted & -0.04\% & 0.01\% & 0.0309 & 0.05\% & 0.51 & 1.70\% & 93.61\% & 66.80\% & 0.00\% \\ 
        ~ & ~ & Elasticnet & -0.25\% & 0.19\% & 0.0458 & 0.44\% & 0.43 & 3.69\% & 94.38\% & 57.50\% & 29.61\% \\ 
        ~ & ~ & Lasso & -0.15\% & 0.18\% & 0.0429 & 0.33\% & 0.42 & 8.61\% & 95.28\% & 59.83\% & 23.33\% \\ 
        ~ & ~ & Neural network & 0.90\% & -1.21\% & -0.0830 & -2.11\% & 0.39 & 15.85\% & 96.66\% & 72.83\% & -6.28\% \\ 
        ~ & ~ & Support vector machine & 2.93\% & 0.79\% & -0.1137 & -2.14\% & 0.36 & 2.02\% & 94.41\% & 65.23\% & 18.63\% \\ 
        ~ & ~ & Super learner & 1.85\% & 0.61\% & -0.0626 & -1.23\% & 0.36 & 3.21\% & 94.47\% & 59.09\% & 31.23\% \\   \cline{2-12}
        
        ~ & \multirow{6}{*}{1.3} & Unadjusted & 0.01\% & 0.08\% & 0.0178 & 0.07\% & 0.53 & 3.64\% & 94.04\% & 90.69\% & 0.00\% \\ 
        ~ & ~ & Elasticnet & -0.14\% & 0.25\% & 0.0322 & 0.39\% & 0.44 & 5.68\% & 94.50\% & 89.42\% & 33.23\% \\ 
        ~ & ~ & Lasso & -0.05\% & 0.33\% & 0.0328 & 0.38\% & 0.44 & 10.10\% & 95.43\% & 90.65\% & 29.44\% \\ 
        ~ & ~ & Neural network & 0.62\% & -0.77\% & -0.0703 & -1.39\% & 0.39 & 15.94\% & 96.36\% & 94.84\% & -50.62\% \\ 
        ~ & ~ & Support vector machine & 1.98\% & 1.11\% & -0.0508 & -0.87\% & 0.34 & 8.84\% & 95.53\% & 92.56\% & 16.34\% \\ 
        ~ & ~ & Super learner & 1.55\% & 1.07\% & -0.0289 & -0.48\% & 0.36 & 5.96\% & 94.89\% & 90.79\% & 28.11\% \\  \cline{2-12}
        
        ~ & \multirow{6}{*}{1.0} & Unadjusted & 0.08\% & -0.09\% & -0.0095 & -0.17\% & 0.57 & 2.88\% & 94.33\% & 5.68\% & \multicolumn{1}{c}{-} \\ 
        ~ & ~ & Elasticnet & -0.10\% & 0.10\% & 0.0112 & 0.20\% & 0.46 & 6.69\% & 94.62\% & 5.38\% & \multicolumn{1}{c}{-} \\ 
        ~ & ~ & Lasso & -0.01\% & 0.22\% & 0.0145 & 0.23\% & 0.46 & 10.73\% & 95.38\% & 4.62\% & \multicolumn{1}{c}{-} \\ 
        ~ & ~ & Neural network & 0.43\% & -0.54\% & -0.0629 & -0.97\% & 0.39 & 17.54\% & 96.58\% & 3.42\% & \multicolumn{1}{c}{-} \\ 
        ~ & ~ & Support vector machine & 1.35\% & 1.23\% & -0.0067 & -0.13\% & 0.35 & 9.66\% & 96.10\% & 3.90\% & \multicolumn{1}{c}{-} \\ 
        ~ & ~ & Super learner & 1.36\% & 1.23\% & -0.0096 & -0.13\% & 0.37 & 7.00\% & 95.36\% & 4.64\% & \multicolumn{1}{c}{-} \\    \hline

         \multirow{18}{*}{$n = 60$} & \multirow{6}{*}{1.9} & Unadjusted & 0.11\% & 0.15\% & 0.1520 & 0.05\% & 1.02 & 13.52\% & 94.01\% & 77.54\% & 0.00\% \\ 
        ~ & ~ & Elasticnet & -0.20\% & 0.60\% & 0.1314 & 0.80\% & 0.73 & 10.54\% & 93.90\% & 71.51\% & 49.44\% \\ 
        ~ & ~ & Lasso & -0.09\% & 0.54\% & 0.1293 & 0.63\% & 0.72 & 15.18\% & 94.65\% & 73.80\% & 45.65\% \\ 
        ~ & ~ & Neural network & 2.68\% & -1.34\% & -0.1937 & -4.02\% & 0.48 & 7.05\% & 94.12\% & 83.01\% & 44.21\% \\ 
        ~ & ~ & Support vector machine & 3.72\% & 0.02\% & -0.1939 & -3.70\% & 0.44 & -1.60\% & 93.09\% & 77.60\% & 59.37\% \\ 
        ~ & ~ & Super learner & 2.81\% & 0.85\% & -0.1078 & -1.97\% & 0.46 & 4.41\% & 94.25\% & 75.02\% & 64.67\% \\  \cline{2-12}
        
        ~ & \multirow{6}{*}{1.3} & Unadjusted & 0.50\% & 0.62\% & 0.0820 & 0.11\% & 1.05 & 20.69\% & 93.91\% & 91.80\% & 0.00\% \\ 
        ~ & ~ & Elasticnet & 0.33\% & 1.01\% & 0.0806 & 0.68\% & 0.75 & 14.72\% & 93.78\% & 90.49\% & 53.62\% \\ 
        ~ & ~ & Lasso & 0.44\% & 1.14\% & 0.0854 & 0.70\% & 0.74 & 20.85\% & 94.64\% & 91.73\% & 51.02\% \\ 
        ~ & ~ & Neural network & 2.37\% & 0.01\% & -0.1357 & -2.36\% & 0.46 & 9.49\% & 93.92\% & 95.12\% & -17.57\% \\ 
        ~ & ~ & Support vector machine & 2.56\% & 1.30\% & -0.0762 & -1.25\% & 0.41 & 7.31\% & 95.04\% & 93.25\% & 59.22\% \\ 
        ~ & ~ & Super learner & 2.81\% & 2.25\% & -0.0463 & -0.55\% & 0.45 & 7.20\% & 95.05\% & 92.72\% & 62.35\% \\   \cline{2-12}
        
        ~ & \multirow{6}{*}{1.0} & Unadjusted & 0.51\% & 0.58\% & 0.0020 & 0.07\% & 1.12 & 33.62\% & 94.08\% & 5.94\% & \multicolumn{1}{c}{-} \\ 
        ~ & ~ & Elasticnet & 0.36\% & 0.92\% & 0.0389 & 0.56\% & 0.79 & 20.74\% & 94.00\% & 6.01\% & \multicolumn{1}{c}{-} \\ 
        ~ & ~ & Lasso & 0.47\% & 1.11\% & 0.0468 & 0.63\% & 0.79 & 25.08\% & 94.53\% & 5.47\% & \multicolumn{1}{c}{-} \\ 
        ~ & ~ & Neural network & 2.02\% & 0.55\% & -0.0953 & -1.47\% & 0.45 & 13.77\% & 94.92\% & 5.08\% & \multicolumn{1}{c}{-} \\ 
        ~ & ~ & Support vector machine & 1.66\% & 1.66\% & 0.0001 & 0.00\% & 0.40 & 10.37\% & 95.28\% & 4.72\% & \multicolumn{1}{c}{-} \\ 
        ~ & ~ & Super learner & 2.58\% & 2.53\% & -0.0044 & -0.04\% & 0.46 & 8.85\% & 95.24\% & 4.78\% & \multicolumn{1}{c}{-} \\  
\bottomrule
\end{tabular*}
\begin{tablenotes}
\item[$^{\rm *}$] For $mOR=1.9$, the estimated asymptotic values were $\pi_0 = 0.2398$,  $\pi_1 = 0.3772$,  $\log mOR = 0.6618$,  and $\pi_1$ - $\pi_0 = 0.1374$.  For $mOR=1.3$,  the estimated asymptotic values were $\pi_0 = 0.2399$,  $\pi_1 = 0.2871$,  $\log mOR = 0.2479$,  and $\pi_1$ - $\pi_0 = 0.0472$. For $mOR=1.0$, the estimated asymptotic values were $\pi_0 = 0.2399$,  $\pi_1 = 0.2399$,  $\log mOR = -1.2\times 10^{-4}$,  and  $\pi_1$ - $\pi_0 = -1.6\times 10^{-5}$. Theoretical AUC of the outcome model in the simulations: 0.8835.
\item[$^{\rm \dagger}$]Type II error (i.e., 100 - power) for $mOR>1.0$, and type I error for $mOR=1.0$.\\
Abbreviations:  $mOR$, marginal odds ratio; RMSE, root mean square error; VEB, variance estimation bias; RSS,  reduction in sample size.
\end{tablenotes}
\end{table*}
\end{center}

\clearpage

\newpage
\subsection*{Additional details of the models used in the simulations}\label{app1}

\begin{center}
\begin{table*}[!ht]%
\scriptsize
\caption{Models used for simulations in the complex scenario as summarized in Figure~\ref{fig1}.\label{tabS1}}
\begin{tabular*}{\textwidth}{@{\extracolsep\fill}ccc@{}}
\toprule
\textbf{Variable} &  \textbf{Role in the study}  & \textbf{Distribution}$^{\tnote{\bf *}}$   \\ \hline
$X_1$  & Continuous   covariate  & $\mathcal{N} (0,1)$           \\
$X_2$  & Continuous    covariate &      $\mathcal{N} (\beta_0 + \beta_1 \times X_1, 1)$               \\
$X_3$   & Continuous  covariate  &    $\mathcal{N} (\beta_0  - \beta_1  \times X_1 -  \beta_2  \times X_2 , 1)$               \\
$X_4$  & Continuous   covariate &        $\mathcal{N} (0,1)$               \\
$X_5$  & Binary   covariate  &               $\mathbbm{1} \big \{ \mathcal{N} (0,1)> 0.67 \big \}$  (i.e., prevalence $\sim$ 25\%)              \\
$X_6$  & Binary   covariate  &       $\mathbbm{1} \big \{ \mathcal{N} (\beta_0 - \beta_1 \times X_4, 1)> -0.40 \big \}$  (i.e., prevalence $\sim$  50\%)         \\
$X_7$ & Continuous   covariate &          $\mathcal{N} (\beta_0 - \beta_1 \times X_5,  1)$         \\
$X_8$  & Binary  covariate    &         $\mathbbm{1} \big \{  \mathcal{N} (\beta_0 + \beta_1 \times X_6,  1)> -0.80)$  (i.e., prevalence $\sim$  75\%)    \\
$X_9$  & Continuous   covariate &   $\mathcal{N} (\beta_0 + \beta_1 \times X_7, 1)$  \\
$X_{10}$  & Continuous  covariate   &  $\mathcal{N} (0,1)$ \\
$X_{11}$  & Binary  covariate   & $\mathbbm{1}  \big \{  \mathcal{N} (\beta_0 + \beta_1 \times X_8, 1)> 0.84 \big \}$ (i.e., prevalence $\sim$  25\%)  \\
$X_{12}$ & Continuous  covariate &  $\mathcal{N} (\beta_0  - \beta_1  \times X_{11} -  \beta_2  \times X_{10} , 1)$ \\
$X_{13}$ & Continuous  covariate   &  $\mathcal{N} (\beta_0 - \beta_1 \times X_{11}, 1)$ \\
$X_{14}$ & Continuous  covariate   &    $\mathcal{N} (0,1)$ \\
$X_{15}$ & Binary  covariate   & $\mathbbm{1} \big \{ \mathcal{N} (0,1)> 0.67 \big \}$ (i.e., prevalence $\sim$ 25\%)  \\
$X_{16}$ & Binary  covariate &  $\mathbbm{1} \big \{ \mathcal{N} (0,1)> 0.67 \big \}$  (i.e., prevalence $\sim$ 25\%)  \\
$X_{17}$ & Continuous  covariate   &    $\mathcal{N} (0,1)$    \\ \hline
$A$ & Binary treatment arm   &   $\mathbbm{1} \big \{ \mathcal{N} (0,1)> 0 \big \}$  (i.e., a 1:1 randomized clinical trial)   \\ \hline
\multirow{2}{*}{$Y$} & \multirow{2}{*}{Binary  outcome} &   $\mathcal{B} \big ( n, p = $ logistic$ \big(  \beta_2 + \beta_3  \times \mathbbm{1} \big \{ X_{2} > -0.44  \}  -   \beta_3  * X_3 + ( \beta_3 / 2) \times X_3^2  + \beta_3 \times X_5 +  \beta_3 \times X_6 +  \beta_3 \times X_9$ \\
&  &  $ +   (\beta_3/2) \times X_{10}^2 -   \beta_3 \times X_{12} -  \beta_3 \times (X_{13} > -0.55) +  \beta_3 \times X_{14} +   \beta_3 \times X_{15} +   (\beta_3/2) \times  A \times X_{14} +   \beta_4 \times A    \big )  \big )$      \\ 
\bottomrule
\end{tabular*}
\begin{tablenotes}
\item[$^{\rm *}$] $\mathbbm{1} \big \{  a \} = 1$ if the condition $a$ is true and 0 otherwise;  $\mathcal{N} (\mu,\sigma)$ represents a Gaussian distribution with mean at $\mu$ and standard deviation at $\sigma$; $\mathcal{B} (n,p)$ represents a Binomial distribution with a size $n$ and probability of success  $p$.  The regression coefficients were: $\beta_0 = -0.4 $,   $\beta_1 = log(2)$,  $\beta_2 = -2$,  $\beta_3 = log(2)$ and $\beta_4 = log(3) ; log(1.5) ; log(0.9729)$ to obtain mOR values of 1.9 ; 1.3 ; 1.0 respectively.
\end{tablenotes}
\end{table*}
\end{center}

\begin{center}
\begin{table*}[!ht]%
\scriptsize
\caption{Models used for simulations in the simple scenario as summarized in Figure~\ref{fig2}.\label{tabS2}}
\begin{tabular*}{\textwidth}{@{\extracolsep\fill}ccc@{}}
\toprule
\textbf{Variable} &  \textbf{Role in the study}  & \textbf{Distribution}$^{\tnote{\bf *}}$   \\ \hline
$X_1$  & Continuous   covariate  & $\mathcal{N} (0,1)$           \\
$X_2$  & Continuous    covariate &      $\mathcal{N} (0,1)$                \\
$X_3$   & Continuous  covariate  &      $\mathcal{N} (0,1)$              \\
$X_4$  & Continuous   covariate &     $\mathbbm{1} \big \{ \mathcal{N} (0,1)< -0.67 \big \}$  (i.e., prevalence $\sim$ 25\%)                    \\
$X_5$  & Binary   covariate  &       $\mathbbm{1} \big \{ \mathcal{N} (0,1)< 0 \big \}$  (i.e., prevalence $\sim$ 50\%)         \\
$X_6$  & Binary   covariate  &     $\mathbbm{1} \big \{ \mathcal{N} (0,1)< 0.67 \big \}$  (i.e., prevalence $\sim$ 75\%)       \\  \hline
$A$ & Binary treatment arm   &    $\mathbbm{1} \big \{ \mathcal{N} (0,1)> 0 \big \}$  (i.e., a 1:1 randomized clinical trial)    \\ \hline 
$Y$ &  Binary outcome & $\mathcal{B} \big ( n, p = $ logistic$ \big(  \beta_0 + \beta_1  \times X_1 + \beta_1  \times X_2 + \beta_1  \times X_4 + \beta_1  \times X_5 +   \beta_2 \times A    \big )  \big )$   \\
\bottomrule
\end{tabular*}
\begin{tablenotes}
\item[$^{\rm *}$] $\mathbbm{1} \big \{  a \} = 1$ if the condition $a$ is true and 0 otherwise;  $\mathcal{N} (\mu,\sigma)$ represents a Gaussian distribution with mean at $\mu$ and standard deviation at $\sigma$; $\mathcal{B} (n,p)$ represents a Binomial distribution with a size $n$ and probability of success  $p$.  The regression coefficients were: $\beta_0 = -3 $,   $\beta_1 = log(4)$  and $\beta_2 = log(3) ; log(1.5) ; log(1)$ to obtain mOR values of 1.9 ; 1.3 ; 1.0 respectively. For the simple scenario with reduced predicative performance the regression coefficients were: $\beta_0 = -0.6 $,   $\beta_1 = 0.2$  and $\beta_2 = log(3) ; log(1.5) ; log(1)$ to obtain mOR values of 3.0 ; 1.5 ; 1.0 respectively.
\end{tablenotes}
\end{table*}
\end{center}

\newpage
\subsection*{Definitions of the criteria used to evaluate the performance of the methods}\label{app2}

Let $\hat{\theta}_s$ and $\sigma(\hat{\theta}_s)$ be the marginal treatment effect and the related standard deviation both estimated in the $s$th simulated data set ($s=1, ..., S$) and $\theta$ the true value. Let $\hat{\theta}_{unadjusted,s}$ and $\sigma(\hat{\theta}_{unadjusted,s})$ be the unadjusted treatment effect and the related standard deviation in the $s$th simulated data set ($s=1, ..., S$). To define the upper and lower bounds of the confidence intervals, let \( Q_{0.025}(\hat{\theta}_s) \) be the \(2.5\%\) quantile of \( \hat{\theta}_s \) and \( Q_{0.975}(\hat{\theta}_s) \) be the \(97.5\%\) quantile of \( \hat{\theta}_s \), respectively. 
The criteria we used were:
  \begin{itemize}[itemsep=16pt]
 \item Mean bias: $  \left[ \big ( \sum_s^{S} (\hat{\theta}_s - \theta)   \big )   \big / S \right] \times 100$ 
  \item Variance estimation bias:  $\left[ \left( S^{-1} \sum_{s=1}^{S} \sigma(\hat{\theta}_s)  - \sqrt{\frac{\sum_{s=1}^{S} (\hat{\theta}_s - \theta)^2}{S-1}} \right) \big / \sqrt{\frac{\sum_{s=1}^{S} (\hat{\theta}_s - \theta)^2}{S-1}} \right] \times 100$
   \item Root mean square error:  $ \sqrt{S^{-1} \sum_{s=1}^{S} (\hat{\theta}_s - \theta)^2}$
    \item Empirical coverage rate of the nominal 95\% confidence interval:  $S^{-1} \sum_{s=1}^{S} \mathbbm{1}(\theta \in [Q_{0.025}(\hat{\theta}_s), Q_{0.975}(\hat{\theta}_s)])$
     \item Type I Error: $S^{-1} \sum_{s=1}^{S} \mathbbm{1}\left[Q_{0.025}(\hat{\theta}_s) > 0 \cup Q_{0.975}(\hat{\theta}_s) < 0\right] \times 100$
     \item Type II Error: $100 - \left(S^{-1} \sum_{s=1}^{S} \mathbbm{1}\left[Q_{0.025}(\hat{\theta}_s) > 0 \cup Q_{0.975}(\hat{\theta}_s) < 0\right] \times 100 \right)$
      \item Reduction in sample size: $\left(1 - \left(\frac{z_{\text{ref}}}{z_{\text{adj}}}\right)^2\right) \times 100$ where $z_{\text{ref}} = \frac{\sum_{s=1}^{S} \hat{\theta}_{\text{unadjusted}, s}}{\sum_{s=1}^{S} \hat{s}(\hat{\theta}_{\text{unadjusted}, s})}
\quad \text{and} \quad
z_{\text{adj}} = \frac{\sum_{s=1}^{S} \hat{\theta}_s}{\sum_{s=1}^{S} \hat{s}(\hat{\theta}_s)}$

      \item Area under the curve: $\frac{1}{S^2} \sum_{s=1}^{S} \sum_{l=1}^{S} \mathbbm{1}\left[\mathds{P}(Y_s=1 \mid X_s) > \mathds{P}(Y_l=1 \mid X_l)\right]$

 \end{itemize}

\newpage

\subsection*{Additional result tables and figures of the simulations and applications}\label{app3}

\begin{center}
\begin{table*}[!ht]%
\scriptsize
\caption{Results of the simulations for the simple scenario with reduced predictive performance, ie. lower regression coefficients.\label{tabs3}}
\begin{tabular*}{\textwidth}{@{\extracolsep\fill}c|c|c|rrrr|rrrrr}
\toprule
\multirow{2}{*}{\makecell{\textbf{Sample} \\ \textbf{size}}}  &  \multirow{2}{*}{\textbf{mOR}\tnote{$^*$} } & \multirow{2}{*}{\textbf{Method} } & \multicolumn{4}{c|}{\textbf{Mean bias}} & \multicolumn{5}{c}{\boldmath$\log\hat{mOR}$} \\\cline{4-7}\cline{8-12}
         &  &  & \boldmath$\hat{\pi}_0$ & \boldmath$\hat{\pi}_1$ &  \boldmath$\log\hat{mOR}$ &\boldmath$\hat{\Delta}$ & \textbf{RMSE} & \textbf{VEB} & \textbf{Coverage} & \textbf{Error}\tnote{$^\dagger$} & \textbf{RSS } \\ \hline
        \multirow{18}{*}{$n= 200$}& \multirow{6}{*}{3.0} & Unadjusted & -0.05\% & 0.01\% & 0.0145 & 0.05\% & 0.30 & -0.16\% & 94.57\% & 3.50\% & 0.00\% \\ 
        ~ & ~ & Elasticnet & -0.04\% & -0.02\% & 0.0136 & 0.03\% & 0.30 & 0.41\% & 94.80\% & 3.75\% & -0.87\% \\ 
        ~ & ~ & Lasso & -0.03\% & -0.05\% & 0.0120 & -0.02\% & 0.30 & 1.24\% & 95.08\% & 4.08\% & -2.80\% \\ 
        ~ & ~ & Neural network & -0.07\% & 0.05\% & 0.0292 & 0.12\% & 0.32 & 17.17\% & 97.21\% & 9.88\% & -53.87\% \\ 
        ~ & ~ & Support vector machine & 0.76\% & -1.82\% & -0.0960 & -2.58\% & 0.42 & -12.38\% & 85.94\% & 11.81\% & -88.16\% \\ 
        ~ & ~ & Super learner & -0.10\% & 0.13\% & -0.0314 & 0.60\% & 0.38 & 13.05\% & 89.69\% & 20.58\% & -119.71\% \\      \cline{2-12}
        
        ~ & \multirow{6}{*}{1.5} & Unadjusted & 0.13\% & -0.09\% & -0.0046 & -0.21\% & 0.29 & -0.33\% & 94.32\% & 72.64\% & 0.00\% \\ 
        ~ & ~ & Elasticnet & 0.15\% & -0.04\% & -0.0034 & -0.19\% & 0.29 & 2.02\% & 94.74\% & 72.95\% & -4.50\% \\ 
        ~ & ~ & Lasso & 0.16\% & -0.04\% & -0.0039 & -0.20\% & 0.29 & 3.04\% & 94.77\% & 73.78\% & -6.82\% \\ 
        ~ & ~ & Neural network & 0.06\% & -0.20\% & -0.0019 & -0.26\% & 0.31 & 14.52\% & 97.10\% & 83.10\% & -48.41\% \\ 
        ~ & ~ & Support vector machine & 2.33\% & -1.65\% & -0.1626 & -3.98\% & 0.29 & -19.00\% & 86.59\% & 84.36\% & -83.86\% \\ 
        ~ & ~ & Super learner & 3.19\% & -0.44\% & -0.1134 & -3.63\% & 0.30 & 37.77\% & 96.78\% & 91.83\% & -269.05\% \\      \cline{2-12}
        
        ~ & \multirow{6}{*}{1.0} & Unadjusted & 0.07\% & 0.00\% & -0.0027 & -0.06\% & 0.30 & -0.52\% & 94.36\% & 5.64\% & \multicolumn{1}{c}{-} \\ 
        ~ & ~ & Elasticnet & 0.15\% & 0.12\% & -0.0015 & -0.03\% & 0.30 & 2.86\% & 94.59\% & 5.40\% & \multicolumn{1}{c}{-} \\ 
        ~ & ~ & Lasso & 0.16\% & 0.14\% & -0.0012 & -0.02\% & 0.30 & 4.17\% & 94.93\% & 5.07\% & \multicolumn{1}{c}{-} \\ 
        ~ & ~ & Neural network & 0.02\% & -0.14\% & -0.0069 & -0.16\% & 0.32 & 14.02\% & 97.19\% & 2.80\% & \multicolumn{1}{c}{-} \\ 
        ~ & ~ & Support vector machine & 0.29\% & 0.24\% & -0.0022 & -0.05\% & 0.17 & 22.14\% & 98.01\% & 1.98\% & \multicolumn{1}{c}{-} \\ 
        ~ & ~ & Super learner & 0.28\% & 0.03\% & -0.0025 & -0.10\% & 0.21 & 88.88\% & 99.17\% & 0.83\% & \multicolumn{1}{c}{-} \\         \hline

         \multirow{18}{*}{$n = 100$} & \multirow{6}{*}{3.0} & Unadjusted & 0.05\% & 0.05\% & 0.0270 & 0.00\% & 0.44 & -0.53\% & 94.06\% & 23.99\% & 0.00\% \\ 
        ~ & ~ & Elasticnet & 0.13\% & -0.12\% & 0.0180 & -0.24\% & 0.44 & 1.11\% & 94.66\% & 25.98\% & -6.07\% \\ 
        ~ & ~ & Lasso & 0.15\% & -0.22\% & 0.0133 & -0.37\% & 0.44 & 2.65\% & 95.05\% & 27.72\% & -10.49\% \\ 
        ~ & ~ & Neural network & 0.51\% & -0.32\% & 0.0102 & -0.82\% & 0.46 & 14.13\% & 96.97\% & 37.88\% & -47.03\% \\ 
        ~ & ~ & Support vector machine & 3.73\% & -4.23\% & -0.3217 & -7.96\% & 0.58 & -27.61\% & 73.17\% & 45.85\% & -94.75\% \\ 
        ~ & ~ & Super learner & 1.04\% & -1.28\% & -0.1393 & -2.32\% & 0.49 & 2.02\% & 93.21\% & 46.64\% & -78.94\% \\       \cline{2-12}
        
        ~ & \multirow{6}{*}{1.5} & Unadjusted & -0.06\% & -0.12\% & 0.0070 & -0.06\% & 0.42 & -0.09\% & 94.28\% & 82.80\% & 0.00\% \\ 
        ~ & ~ & Elasticnet & 0.14\% & 0.03\% & 0.0055 & -0.12\% & 0.43 & 6.27\% & 94.50\% & 83.51\% & -15.60\% \\ 
        ~ & ~ & Lasso & 0.20\% & 0.04\% & 0.0039 & -0.16\% & 0.43 & 8.51\% & 94.91\% & 84.26\% & -21.60\% \\ 
        ~ & ~ & Neural network & 0.09\% & -0.26\% & 0.0021 & -0.35\% & 0.44 & 12.14\% & 96.44\% & 88.86\% & -39.91\% \\ 
        ~ & ~ & Support vector machine & 2.43\% & -2.06\% & -0.1835 & -4.48\% & 0.34 & -7.76\% & 92.68\% & 91.03\% & -93.79\% \\ 
        ~ & ~ & Super learner & 1.21\% & -0.82\% & -0.0931 & -2.04\% & 0.37 & 33.86\% & 97.38\% & 92.39\% & -132.07\% \\    \cline{2-12}
        
        ~ & \multirow{6}{*}{1.0} & Unadjusted & 0.04\% & 0.06\% & 0.0004 & 0.02\% & 0.43 & -1.16\% & 94.30\% & 5.68\% & \multicolumn{1}{c}{-} \\ 
        ~ & ~ & Elasticnet & 0.34\% & 0.49\% & 0.0057 & 0.15\% & 0.44 & 7.36\% & 94.61\% & 5.39\% & \multicolumn{1}{c}{-} \\ 
        ~ & ~ & Lasso & 0.41\% & 0.56\% & 0.0061 & 0.16\% & 0.44 & 9.70\% & 94.83\% & 5.15\% & \multicolumn{1}{c}{-} \\ 
        ~ & ~ & Neural network & 0.07\% & 0.03\% & -0.0023 & -0.04\% & 0.45 & 11.40\% & 96.72\% & 3.29\% & \multicolumn{1}{c}{-} \\ 
        ~ & ~ & Support vector machine & 0.31\% & 0.32\% & 0.0006 & 0.02\% & 0.25 & 21.23\% & 97.41\% & 2.59\% & \multicolumn{1}{c}{-} \\ 
        ~ & ~ & Super learner & 0.72\% & -0.06\% & 0.0001 & -0.77\% & 0.33 & 48.43\% & 98.18\% & 1.82\% & \multicolumn{1}{c}{-} \\        \hline

         \multirow{18}{*}{$n = 60$} & \multirow{7}{*}{3.0} & Unadjusted & -0.11\% & -0.06\% & 0.0715 & 0.05\% & 0.68 & 3.12\% & 94.40\% & 45.08\% & 0.00\% \\ 
        ~ & ~ & Elasticnet & 0.04\% & -0.47\% & 0.0400 & -0.51\% & 0.62 & 5.44\% & 94.98\% & 48.16\% & 7.81\% \\ 
        ~ & ~ & Lasso & 0.07\% & -0.63\% & 0.0328 & -0.71\% & 0.62 & 7.32\% & 95.45\% & 49.97\% & 2.97\% \\
        ~ & ~ & Neural network & 1.54\% & -1.49\% & -0.0626 & -3.03\% & 0.58 & 13.01\% & 96.66\% & 58.73\% & -10.12\% \\ 
        ~ & ~ & Support vector machine & 5.19\% & -5.72\% & -0.4517 & -10.91\% & 0.67 & -28.94\% & 78.41\% & 65.74\% & -44.80\% \\ 
        ~ & ~ & Super learner & 1.99\% & -2.34\% & -0.1683 & -4.33\% & 0.56 & 7.36\% & 94.75\% & 62.49\% & -16.09\% \\      \cline{2-12}
        
        ~ & \multirow{7}{*}{1.5} & Unadjusted & -0.03\% & 0.00\% & 0.0240 & 0.03\% & 0.58 & 4.46\% & 94.41\% & 87.34\% & 0.00\% \\ 
        ~ & ~ & Elasticnet & 0.36\% & 0.24\% & 0.0177 & -0.11\% & 0.58 & 12.01\% & 94.80\% & 87.66\% & -17.84\% \\ 
        ~ & ~ & Lasso & 0.43\% & 0.24\% & 0.0153 & -0.19\% & 0.58 & 14.76\% & 95.10\% & 88.38\% & -25.35\% \\ 
        ~ & ~ & Neural network & 0.72\% & -0.57\% & -0.0310 & -1.28\% & 0.55 & 10.42\% & 95.99\% & 92.12\% & -28.23\% \\ 
        ~ & ~ & Support vector machine & 2.60\% & -2.16\% & -0.1962 & -4.76\% & 0.40 & -0.92\% & 93.29\% & 92.97\% & -71.44\% \\ 
        ~ & ~ & Super learner & 1.21\% & -0.48\% & -0.0774 & -1.69\% & 0.47 & 24.51\% & 96.69\% & 93.33\% & -57.11\% \\       \cline{2-12}
        
        ~ & \multirow{7}{*}{1.0} & Unadjusted & -0.07\% & 0.05\% & 0.0060 & 0.12\% & 0.61 & 6.05\% & 94.01\% & 6.00\% & \multicolumn{1}{c}{-} \\ 
        ~ & ~ & Elasticnet & 0.51\% & 0.91\% & 0.0173 & 0.40\% & 0.61 & 13.61\% & 94.31\% & 5.71\% & \multicolumn{1}{c}{-}  \\ 
        ~ & ~ & Lasso & 0.58\% & 1.02\% & 0.0190 & 0.44\% & 0.61 & 17.00\% & 94.57\% & 5.44\% & \multicolumn{1}{c}{-} \\
        ~ & ~ & Neural network & 0.11\% & 0.04\% & -0.0059 & -0.06\% & 0.55 & 10.18\% & 96.43\% & 3.57\% & \multicolumn{1}{c}{-} \\ 
        ~ & ~ & Support vector machine & 0.28\% & 0.29\% & 0.0001 & 0.01\% & 0.33 & 19.32\% & 96.90\% & 3.11\% & \multicolumn{1}{c}{-} \\ 
        ~ & ~ & Super learner & 0.47\% & 0.76\% & 0.0009 & 0.29\% & 0.47 & 27.52\% & 97.05\% & 2.95\% & \multicolumn{1}{c}{-} \\  
\bottomrule
\end{tabular*}
\begin{tablenotes}
\item[$^{\rm *}$] For $mOR=3.0$, the estimated asymptotic values were $\pi_0 = 0.3918$,  $\pi_1 = 0.6534 $,  $\log mOR = 1.085$,  and $\pi_1$ - $\pi_0 = 0.2616$.
For $mOR=1.5$,  the estimated asymptotic values were $\pi_0 = 0.3917$,  $\pi_1 = 0.4891$,  $\log mOR = 0.4120$,  and $\pi_1$ - $\pi_0 = 0.0974$.
For $mOR=1.0$, the estimated asymptotic values were $\pi_0 = 0.3918$,  $\pi_1 = 0.3918$,  $\log mOR = -1.9\times 10^{-4}$,  and  $\pi_1$ - $\pi_0 = -4.6\times 10^{-5}$. Theoretical AUC of the outcome model in the simulations: 0.6714.
\item[$^{\rm \dagger}$]Type II error (i.e., 100 - power) for $mOR>1.0$, and type I error for $mOR=1.0$.\\
Abbreviations:  $mOR$, marginal odds ratio; RMSE, root mean square error; VEB, variance estimation bias; RSS,  reduction in sample size.
\end{tablenotes}
\end{table*}
\end{center}

\begin{center}
\begin{table*}[!ht]%
\scriptsize
\caption{Results of the simulations for the simple scenario with reduced predictive performance, ie. lower regression coefficients without considering interactions or B-splines for the penalized methods.\label{tabs3b}}
\begin{tabular*}{\textwidth}{@{\extracolsep\fill}c|c|c|rrrr|rrrrr}
\toprule
\multirow{2}{*}{\makecell{\textbf{Sample} \\ \textbf{size}}}  &  \multirow{2}{*}{\textbf{mOR}\tnote{$^*$} } & \multirow{2}{*}{\textbf{Method} } & \multicolumn{4}{c|}{\textbf{Mean bias}} & \multicolumn{5}{c}{\boldmath$\log\hat{mOR}$} \\\cline{4-7}\cline{8-12}
         &  &  & \boldmath$\hat{\pi}_0$ & \boldmath$\hat{\pi}_1$ &  \boldmath$\log\hat{mOR}$ &\boldmath$\hat{\Delta}$ & \textbf{RMSE} & \textbf{VEB} & \textbf{Coverage} & \textbf{Error}\tnote{$^\dagger$} & \textbf{RSS } \\ \hline
        \multirow{9}{*}{$n= 200$}& \multirow{3}{*}{3.0} & Unadjusted & -0.05\% & -0.00\% & 0.0141 & 0.05\% & 0.30 & -0.09\% & 94.54\% & 3.42\% & 0.00\% \\ 
        ~ & ~ & Elasticnet & -0.05\% & -0.00\% & 0.0144 & 0.05\% & 0.30 & -0.54\% & 94.30\% & 3.24\% & 1.54\% \\ 
        ~ & ~ & Lasso & -0.05\% & -0.01\% & 0.0143 & 0.04\% & 0.30 & -0.42\% & 94.35\% & 3.33\% & 1.12\% \\      \cline{2-12}
        
        ~ & \multirow{3}{*}{1.5} & Unadjusted & 0.13\% & -0.09\% & -0.0047 & -0.22\% & 0.29 & -0.31\% & 94.33\% & 72.48\% & 0.00\% \\ 
        ~ & ~ & Elasticnet & 0.11\% & -0.07\% & -0.0032 & -0.18\% & 0.29 & -0.93\% & 94.27\% & 71.49\% & 2.09\% \\ 
        ~ & ~ & Lasso & 0.11\% & -0.07\% & -0.0031 & -0.18\% & 0.29 & -0.91\% & 94.22\% & 71.57\% & 1.89\% \\      \cline{2-12}
        
        ~ & \multirow{3}{*}{1.0} & Unadjusted & 0.06\% & 0.01\% & -0.0026 & -0.06\% & 0.30 & -0.60\% & 94.42\% & 5.56\% & \multicolumn{1}{c}{-} \\ 
        ~ & ~ & Elasticnet & 0.08\% & 0.01\% & -0.0030 & -0.07\% & 0.30 & -1.39\% & 94.23\% & 5.77\% & \multicolumn{1}{c}{-} \\ 
        ~ & ~ & Lasso & 0.09\% & 0.01\% & -0.0031 & -0.07\% & 0.30 & -1.32\% & 94.22\% & 5.76\% & \multicolumn{1}{c}{-} \\         \hline

         \multirow{9}{*}{$n = 100$} & \multirow{3}{*}{3.0} & Unadjusted & 0.05\% & 0.05\% & 0.0272 & 0.01\% & 0.44 & -0.57\% & 94.02\% & 23.85\% & 0.00\% \\ 
        ~ & ~ & Elasticnet & 0.06\% & 0.04\% & 0.0265 & -0.02\% & 0.44 & -1.44\% & 93.92\% & 22.86\% & 1.99\% \\ 
        ~ & ~ & Lasso & 0.08\% & 0.01\% & 0.0246 & -0.08\% & 0.44 & -1.01\% & 94.10\% & 23.39\% & 0.76\% \\      \cline{2-12}
        
        ~ & \multirow{3}{*}{1.5} & Unadjusted & -0.06\% & -0.11\% & 0.0078 & -0.04\% & 0.42 & -0.08\% & 94.08\% & 82.56\% & 0.00\% \\ 
        ~ & ~ & Elasticnet & -0.07\% & -0.08\% & 0.0097 & -0.01\% & 0.43 & -1.45\% & 93.74\% & 82.06\% & 2.41\% \\ 
        ~ & ~ & Lasso & -0.05\% & -0.07\% & 0.0093 & -0.02\% & 0.43 & -1.24\% & 93.93\% & 82.12\% & 1.55\% \\    \cline{2-12}
        
        ~ & \multirow{3}{*}{1.0} & Unadjusted & 0.04\% & 0.05\% & 0.0001 & 0.01\% & 0.43 & -1.12\% & 94.23\% & 5.77\% & \multicolumn{1}{c}{-} \\ 
        ~ & ~ & Elasticnet & 0.08\% & 0.08\% & -0.0004 & 0.00\% & 0.44 & -2.27\% & 93.92\% & 6.08\% & \multicolumn{1}{c}{-} \\ 
        ~ & ~ & Lasso & 0.10\% & 0.09\% & -0.0006 & -0.00\% & 0.44 & -2.11\% & 94.07\% & 5.91\% & \multicolumn{1}{c}{-} \\        \hline

         \multirow{9}{*}{$n = 60$} & \multirow{3}{*}{3.0} & Unadjusted & -0.11\% & -0.05\% & 0.0722 & 0.07\% & 0.68 & 2.93\% & 94.30\% & 44.72\% & 0.00\% \\ 
        ~ & ~ & Elasticnet & -0.12\% & -0.06\% & 0.0613 & 0.06\% & 0.62 & -1.34\% & 93.88\% & 43.85\% & 21.41\% \\ 
        ~ & ~ & Lasso & -0.07\% & -0.13\% & 0.0580 & -0.06\% & 0.63 & -0.62\% & 93.95\% & 44.82\% & 19.54\% \\      \cline{2-12}
        
        ~ & \multirow{3}{*}{1.5} & Unadjusted & -0.04\% & -0.01\% & 0.0238 & 0.03\% & 0.58 & 4.46\% & 94.28\% & 87.34\% & 0.00\% \\ 
        ~ & ~ & Elasticnet & -0.02\% & 0.05\% & 0.0234 & 0.07\% & 0.58 & -1.03\% & 93.82\% & 86.87\% & 10.62\% \\ 
        ~ & ~ & Lasso & 0.02\% & 0.04\% & 0.0220 & 0.02\% & 0.58 & -0.43\% & 93.98\% & 87.06\% & 8.61\% \\        \cline{2-12}
        
        ~ & \multirow{3}{*}{1.0} & Unadjusted & -0.09\% & 0.05\% & 0.0063 & 0.13\% & 0.61 & 5.88\% & 94.17\% & 5.79\% & \multicolumn{1}{c}{-} \\ 
        ~ & ~ & Elasticnet & -0.02\% & 0.13\% & 0.0073 & 0.16\% & 0.60 & -1.53\% & 93.58\% & 6.43\% & \multicolumn{1}{c}{-} \\ 
        ~ & ~ & Lasso & 0.02\% & 0.19\% & 0.0076 & 0.17\% & 0.61 & -1.16\% & 93.60\% & 6.29\% & \multicolumn{1}{c}{-} \\  
\bottomrule
\end{tabular*}
\begin{tablenotes}
\item[$^{\rm *}$] For $mOR=3.0$, the estimated asymptotic values were $\pi_0 = 0.3918$,  $\pi_1 = 0.6534 $,  $\log mOR = 1.085$,  and $\pi_1$ - $\pi_0 = 0.2616$.
For $mOR=1.5$,  the estimated asymptotic values were $\pi_0 = 0.3917$,  $\pi_1 = 0.4891$,  $\log mOR = 0.4120$,  and $\pi_1$ - $\pi_0 = 0.0974$.
For $mOR=1.0$, the estimated asymptotic values were $\pi_0 = 0.3918$,  $\pi_1 = 0.3918$,  $\log mOR = -1.9\times 10^{-4}$,  and  $\pi_1$ - $\pi_0 = -4.6\times 10^{-5}$. Theoretical AUC of the outcome model in the simulations: 0.6714.
\item[$^{\rm \dagger}$]Type II error (i.e., 100 - power) for $mOR>1.0$, and type I error for $mOR=1.0$.\\
Abbreviations:  $mOR$, marginal odds ratio; RMSE, root mean square error; VEB, variance estimation bias; RSS,  reduction in sample size.
\end{tablenotes}
\end{table*}
\end{center}

\clearpage

\begin{center}
\begin{table*}[!ht]%
\scriptsize
\caption{Baseline description of the kidney transplant recipients according to the 2 studied arms\label{tabS3}}
\begin{tabular*}{\textwidth}{@{\extracolsep\fill}lccc@{}}
\toprule
\textbf{} &  \makecell{\textbf{ATG} \\ \textbf{(n=113)}}  & \makecell{\textbf{DAC} \\ \textbf{(n=114)}} & \textbf{SMD}   \\ \hline
\textbf{Age, year (mean ± SD)} & 45.4 ± 10.3 & 46.9 ± 9.0 & 17\% \\
\textbf{Males, n (\%)} & 52 (46.0) & 59 (51.8) & 11\% \\
\textbf{Weight, kg (mean ± SD)} & 65 ± 17 & 65 ± 13 & 2\%  \\
Cause of end stage renal disease, n (\%) & & & \\
\hspace{5mm}\textbf{Glomerulonephritis} & 51 (45.2) & 45 (39.4)  & 12\%  \\
\hspace{5mm}Uropathy & 11 (9.7) & 15 (13.2) & \\
\hspace{5mm}Autosomal dominant polycystic kidney disease & 10 (8.9) & 10 (8.8) & \\
\hspace{5mm}Diabetes & 4 (3.5) & 2 (1.7) & \\
\hspace{5mm}Other & 26 (23.0) & 25 (21.9) & \\
\hspace{5mm}Unknown & 11 (9.7) & 17 (15.0) & \\
\textbf{Number of human leukocyte antigens mismatches (mean ± SD)}  & & & 2\% \\
\hspace{5mm}Human leukocyte antigens A & 0.9 ± 0.7 & 0.9 ± 0.7 & \\
\hspace{5mm}Human leukocyte antigens B & 1.1 ± 0.7 & 1.1 ± 0.8 & \\
\hspace{5mm}Human leukocyte antigens DR & 0.9 ± 0.7 & 0.9 ± 0.8&  \\
\textbf{Number of grafts, n(\%)} & & & 5\% \\
\hspace{5mm} 1 or 2 & 89 (78.8) & 92 (80.7)  & \\
\hspace{5mm} 3 or 4 & 24 (21.2) & 22 (19.3)  & \\
First graft, n (\%): & 30 (26.5) & 34 (29.8) & \\
\hspace{5mm}Current panel reactive antibodies, \% (mean ± SD) & 35 ± 32 & 39 ± 33 & \\
\hspace{5mm}Peak panel reactive antibodies, \% (mean ± SD) & 77 ± 20 & 79 ± 20 & \\
Second graft, n (\%): & 59 (52.2) & 58 (50.9) & \\
\hspace{5mm}Current panel reactive antibodies, \% (mean ± SD) & 35 ± 29 & 39 ± 31 & \\
\hspace{5mm}Peak panel reactive antibodies, \% (mean ± SD) & 69 ± 23 & 75 ± 18 & \\
Third and fourth graft, n (\%): & 24 (21.2) & 22 (19.3) & \\
\hspace{5mm}Current panel reactive antibodies, \% (mean ± SD) & 26 ± 30 & 27 ± 31 & \\
\hspace{5mm}Peak panel reactive antibodies, \% (mean ± SD) & 60±  30 & 61 ± 27 & \\
All patients & & & \\
\hspace{5mm}Current panel reactive antibodies, \% (mean ± SD) & 33 ± 30 & 37 ± 32 & \\
\hspace{5mm}Peak panel reactive antibodies, \% (mean ± SD) & 69 ± 25 & 74 ± 22&  \\
\hspace{5mm}\% with panel reactive antibodies >80\% & 8.80\% & 11.40\% & \\
\textbf{Cold ischemia time, h (mean ± SD)} & 24.0 ± 7.9 & 22.7 ± 6.8 & 20\% \\
Donor  \\
\hspace{5mm}Males, n (\%) & 76 (67.3) & 65 (57.0) & \\
\textbf{\hspace{5mm}Age, year (mean ± SD)} & 44.3 ± 13.8 & 44.6 ± 14.8 & 3\% \\
\textbf{\hspace{5mm}Death from stroke, n (\%)} & 56 (49.6) & 46 (40.4) & 19\% \\
\textbf{Cytomegalovirus serologic status, n (\%)} & & & \\
\hspace{5mm}Donor+Recipient+ & 37 (32.7) & 45 (39.5) & 14\%  \\
\hspace{5mm}Donor+Recipient- & 16 (14.2) & 12 (10.5) & 11\% \\
\hspace{5mm}Donor-Recipient+ & 46 (40.7) & 42 (36.8) & 8\% \\
\hspace{5mm}Donor-Recipient- & 14 (12.4) & 15 (13.2) & 2\% \\
\bottomrule
\end{tabular*}
\begin{tablenotes}
\item[Abbreviations:] DAC, daclizumab; ATG, antithymocyte globulin; SMD, standardized mean difference.
\end{tablenotes}
\end{table*}
\end{center}

\clearpage

\begin{center}
\begin{table*}[!ht]%
\scriptsize
\caption{Baseline description of patients with acute hypoxemic respiratory failure according to the 3 studied arms\label{tabS5}}
\begin{tabular*}{\textwidth}{@{\extracolsep\fill}lccccc@{}}
\toprule
\textbf{} &  \makecell{\textbf{High-Flow} \\ \textbf{Oxygen}  \\ \textbf{(n=106)}}  & \makecell{\textbf{Standard} \\ \textbf{Oxygen} \\ \textbf{(n=94)}}  & \makecell{\textbf{Noninvasive} \\ \textbf{Ventilation} \\ \textbf{(n=110)}} & \makecell{\textbf{SMD} \\ \textbf{Standard} \\ \textbf{Oxygen vs.}  \\ \textbf{High-Flow Oxygen} }  & \makecell{\textbf{SMD} \\ \textbf{Noninvasive} \\ \textbf{Ventilation vs.}  \\ \textbf{High-Flow Oxygen}}   \\ \hline
Age, year (mean ± SD) & 61 ± 16 & 59 ± 17 & 61 ± 17 & 12\%  & 0\% \\
Males,  n (\%) & 75 (71) & 63 (67) & 74 (67) & 8\%  & 8\% \\
Body-mass index\tnote{$^*$} (mean ± SD) & 25 ± 5 & 26 ± 5 & 26 ± 6 & 6\% & 14\% \\
SAPS II\tnote{$^\dagger$} (mean ± SD) & 25 ± 9 & 24 ± 9 & 27 ± 9 & 19\%  & 13\% \\
Current or past smoking, n (\%) & 34 (32) & 36 (38) & 40 (36) & 13\% & 9\% \\
Reason for acute respiratory failure, n (\%) &  &  &  \\
\hspace{5mm}Community-acquired pneumonia & 71 (67) & 57 (61) & 69 (63) & 13\%  & 9\% \\
\hspace{5mm}Hospital-acquired pneumonia & 12 (11) & 13 (14) & 12 (11) & 8\%  & 1\% \\
\hspace{5mm}Other & 19 (22) & 21 (25) & 21 (26) & 9\% & 11\% \\
Bilateral pulmonary infiltrates, n (\%) & 79 (75) & 80 (85) & 85 (77) & 27\% & 6\% \\
Respiratory rate, breaths/min (mean ± SD) & 33 ± 6 & 32 ± 6 & 33 ± 7 & 10\% & 5\% \\
Heart rate,  beats/min (mean ± SD) & 106 ± 21 & 104 ± 16 & 106 ± 21 & 8\%  & 2\% \\
Arterial pressure, mm Hg (mean ± SD) &  &  &  \\
\hspace{5mm}Systolic & 127 ± 24 & 130 ± 22 & 128 ± 21 & 14\% & 6\% \\
\hspace{5mm}Diastolic & 72 ± 16 & 69 ± 14 & 68 ± 15 & 2\% & 6\% \\
\hspace{5mm}Mean & 87 ± 17 & 89 ± 15 & 86 ± 16 & 9\%  & 5\% \\
Arterial blood gas (mean ± SD) &  &  &  \\
\hspace{5mm}pH & 7.43 ± 0.05 & 7.44 ± 0.06 & 7.43 ± 0.06 & 12\% & 1\%  \\
\hspace{5mm}PaO2, mm Hg & 85 ± 31 & 92 ± 32 & 90 ± 36 & 24\% & 17\%  \\
\hspace{5mm}FiO2\tnote{$^\ddagger$} & 0.62 ± 0.19 & 0.63 ± 0.17 & 0.65 ± 0.15 & 5\% & 16\%  \\
\hspace{5mm}PaO2:FiO2, mm Hg & 157 ± 89 & 161 ± 73 & 149 ± 72 & 5\% & 9\% \\
\hspace{5mm}PaCO2, mm Hg & 36 ± 6 & 35 ± 5 & 34 ± 6 & 21\% & 29\% \\
\hspace{5mm}SpO2, \% & 94 ± 4 & 94 ± 3 & 94 ± 4 & 11\% & 19\% \\
\bottomrule
\end{tabular*}
\begin{tablenotes}
\item[Abbreviations:] SMD, standardized mean difference; FiO2, fraction of inspired oxygen; PaCO2, partial pressure of arterial carbon dioxide; PaO2, partial pressure of arterial oxygen.
\item[$^{\rm *}$]	The body-mass index is the weight in kilograms divided by the square of the height in meters.
\item[$^{\rm \dagger}$]	The Simplified Acute Physiology Score (SAPS) II was calculated from 17 variables at enrollment, information about pre-vious health status, and information obtained at admission. Scores range from 0 to 163, with higher scores indicating more severe disease.
\item[$^{\rm \ddagger}$]	Fio2 was measured in 286 patients and was estimated in the remaining patients as follows: (oxygen flow in liters per minute) × 0.3 + 0.21.
\end{tablenotes}
\end{table*}
\end{center}

\clearpage

\end{document}